\documentclass{elsart}

\usepackage{graphicx}
\usepackage{amsmath,amssymb,amsfonts,bbm}
\usepackage{epsfig,psfrag}

\newcommand{\be}{\begin{equation}}
\newcommand{\ee}{\end{equation}}
\newcommand{\ba}{\begin{array}}
\newcommand{\ea}{\end{array}}
\newcommand{\bqa}{\begin{eqnarray}}
\newcommand{\eqa}{\end{eqnarray}}

\newcommand{\ds}{\displaystyle}
\newcommand{\ie}{{\it i.e. }}

\newcommand{\tr}{\mbox{Tr}}

\newcommand{\bra}[1]{\ensuremath{\langle #1 |}}

\newcommand{\ket}[1]{\ensuremath{| #1 \rangle}}
\newcommand{\prj}[1]{\ensuremath{| #1 \rangle \langle #1 |}}

\newcommand{\ovl}[2]{\ensuremath{\langle #1 | #2 \rangle}}
\newcommand{\matel}[3]{\ensuremath{\langle #1 | #2 | #3 \rangle}}
\newcommand{\trol}[3]{\langle{#1}|{#2}\rangle\otimes|{#3}\rangle}

\newcommand{\un}{{\mathbbm 1}}
\newcommand{\cal}{\mathcal}

\newcommand{\me}{m} 
\newcommand{\sv}{{\cal S}} 
\renewcommand{\sc}{\lambda} 
\newcommand{\sch}{\xi} 
\newcommand{\ev}{\mu} 
\newcommand{\inv}{I} 
\newcommand{\nss}{p} 
\newcommand{\dss}{d} 

\newcommand{\Eq}[1]{Eq.~(\ref{#1})}

\begin{document}

\begin{frontmatter}

\title{Measures and dynamics of entangled states}

\author[pks,cft,ufrj]{Florian Mintert}
\author[pks]{Andr\'e R R Carvalho}
\author[cft]{Marek Ku\'s}
\author[pks]{Andreas Buchleitner}

\address[pks]{
  Max-Planck-Institut f\"ur Physik komplexer Systeme,
  N\"othnitzerstr. 38, D-01187 Dresden, Germany}

\address[cft]{
  Centrum Fizyki Teoretycznej Polskiej Akademii Nauk,
  Aleja Lotnik\'ow 32/46,
  PL-02-668 Warszawa, Poland}

\address[ufrj]{
  Instituto de F\'{\i}sica, Universidade Federal do Rio de Janeiro,
  Caixa Postal 68528, 21945-970 Rio de Janeiro, RJ, Brazil}

\begin{abstract}
We develop an original approach for the quantitative characterisation of the
entanglement properties of, possibly mixed, bi- and multipartite quantum
states of arbitrary finite dimension. Particular emphasis is given to the
derivation of reliable estimates which allow for an efficient evaluation of a
specific entanglement measure, {\em concurrence}, for further implementation
in the monitoring of the time evolution of multipartite entanglement under
incoherent environment coupling. The flexibility of the technical machinery
established here is illustrated by its implementation for different, realistic
experimental scenarios.

\end{abstract}

\end{frontmatter}

\newpage

\tableofcontents

\section{Introduction}
Entanglement is one of the central issues of debate in quantum theory
since the beginning of the last century and certainly a key idea when
it comes to distinguish classical and quantum concepts. Moreover, besides this
fundamental aspect, the interest in entangled states has been recently
renewed because their properties lie at the heart of many potential
applications. Be it in quantum computation~\cite{fey,Deu83},
teleportation~\cite{wot93b} or quantum cryptography~\cite{bb84},
entanglement is viewed as an important resource and, as such, must be
quantified. In addition, great experimental progresses in the production,
manipulation and detection of entangled
states~\cite{wine95,wine98b,wein00,hae,wine00,panPRL01,eiblPRL04,roos,rauschSCI00,zhaoNAT04,pashNAT03,berkleySCI04,roosPRL04} 
require such a quantification to be versatile enough to deal with the
states encountered in actual experiments, which are in general mixed and
typically involve several particles.

The first attempt to discern the non-local  
correlations of measurement results induced by entanglement was formulated with
Bell's inequalities \cite{Bell,CHSH}, which underwent a first experimental
check \cite{Papa} in the sixties. Bell's inequalities are capable of
discriminating correlations due to entanglement against those
described by local hidden variable models \cite{EPR}. Later, also entanglement
criteria 
that use special 
three-partite states, without involving inequalities, were
found~\cite{GHZ} and tested experimentally~\cite{wein00}. Albeit able
to reveal the entangled nature of some quantum states, the above criteria
cannot (and do not intend) to quantify the amount of entanglement carried by a
given state.  

Only recently has the problem of finding a quantity that measures quantum
correlations been studied more intensively~\cite{ben96d,wot96,vidal00}. 
Virtually the entire
state-of-the-art theory of entangled quantum states is based on
so-called {\em entanglement measures}, scalar quantities that quantify 
quantum correlations, and
distinguish them from classical ones. For bipartite pure states such
measures exist and are straightforwardly computable. However, if one aspires to
describe realistic states observed in experiments, it is imperative to
allow for a proper quantification also of mixed states entanglement, 
since there is
no system that could be decoupled perfectly from environmental
influences, and mixing is thus unavoidable.

Although several measures for mixed states have been proposed,
no simple criterion of discriminating classical from quantum correlations
is known so far. All proposed measures that unambiguously fulfill this task 
involve some - generally high
dimensional - optimisation procedure, and hardly allow for an explicit
evaluation in concrete cases.\footnote{Negativity \cite{vidal02a} can
be evaluated algebraically, though does not detect all entangled states.}
Only for
states of smallest possible dimensions, i.e., for bipartite two-level
systems, do a few measures exist that can be evaluated
algebraically~\cite{wot98,tsw01,eng01}, though no such measures are known for
arbitrary states of higher dimensional systems.

Hence, there is some mismatch between the more formal part of entanglement
theory, which seeks for the general characterisation of arbitrary quantum
states, and experimental progresses of the last years, which are leading to
the production of (possibly specific classes of) 
ever more complex entangled states of multipartite quantum
systems. For a theoretical analysis of the latter, we need algebraic as well
as numerical tools to describe {\em static} as well as {\em dynamical}  
properties of
multipartite quantum states in a quantitative manner -- what is a nontrivial
target, due to the rapidly increasing complexity of the problem at hand, as
particle number and effective Hilbert space dimensions are increased. Studying
the dynamics of entanglement requires efficiently evaluable quantifiers
thereof as an undispensible ingredient, ideally for arbitrary system sizes. 

In the present review, we attempt to improve on this mismatch, by responding
precisely to this latter requirement: Starting from the formal algebraic
description of a suitable entanglement measure -- {\em concurrence} --, we
derive a hierarchy of bounds~\cite{flo03} and approximations~\cite{floqp} thereof, which imply
progressively reduced computational efforts for its actual evaluation for
bi- or multipartite, mixed quantum states in arbitrary finite
dimensions. After numerical tests of the tightness of these various estimates, we
implement this novel toolbox, to monitor entanglement dynamics under
experimentally realistic conditions~\cite{floandre}. 

We start out with a brief recollection of the most important concepts of
entanglement theory.

\subsection{Entangled states}
\subsubsection{Bipartite entanglement}
\label{bipent}

A bipartite system is a quantum system that is composed of two
physically distinct subsystems.
It is associated with a Hilbert space ${\cal H}$ that
is given by the tensor product ${\cal H}_1\otimes{\cal H}_2$ of the
predefined factor spaces, each of which describes one subsystem.

For pure states one distinguishes two different kinds of states.
A state $\ket{\Psi}$ is called a {\it product state} or
{\it separable}, if it can be written as a tensor product of subsystem
states, \ie if there are states $\ket{\varphi}$ and $\ket{\phi}$ of
the subsystems such that
\be
\ket{\Psi}=\ket{\varphi}\otimes\ket{\phi}\ .
\label{pure-separable-states}
\ee
Such a state describes a situation analogous to a classical one insofar
as the system state contains exactly the information that
is contained in the subsystem states.
A state reduction~\cite{schwabl} caused by a measurement performed on one
subsystem has no influence on the state of the other subsystem.
This means that measurement results on the different subsystems are
uncorrelated (or independent).
In contrast to this, they are correlated for {\it entangled} states,
\ie states that cannot be written as a product of subsystem
states as in \Eq{pure-separable-states}.
Here, a local measurement causes a state reduction of the entire,
\ie bipartite, system state, and therefore changes the
probabilities for potential future measurements
on either subsystem.

For mixed states the situation is more complicated.
Product and separable states are not synonymous anymore.
Whereas the former can be expressed as a tensor product of subsystem
states, \ie
there are states $\varrho_{1}$ and $\varrho_{2}$ describing the
single subsystems, such that
\be
\varrho=\varrho_{1}\otimes\varrho_{2}\ ,
\ee
a convex sum of such product states is needed to represent a general
separable state
\be
\varrho=\sum_i\ p_i\ \varrho^i_{1}\otimes\varrho^i_{2}\ ,
\label{mixed-separable-states}
\ee
where convexity implies positive coefficients $p_i$ that sum up
to unity, $\sum_i\ p_i=1$.
Such a state refers to a situation where
correlations between different subsystems are due to incomplete
knowledge about the system state.
They are characterised completely by the classical probabilities $p_i$.

Quantum correlations, \ie, entanglement, need now be distinguished from
classical correlations - a problem which we will focus on throughout
the largest part of this paper.
Formally, and in a rather non-constructive way, an entangled mixed
state is defined through the non-existence~\cite{wern89} of a convex
decomposition alike \Eq{mixed-separable-states}.
\be
\varrho\neq\sum_i\ p_i\ \varrho^i_{1}\otimes\varrho^i_{2}\, ,
\label{mixed-entangled-states}
\ee
what is the mixed state generalization of the negation of
(\ref{pure-separable-states}).
The correlations contained in such states cannot be characterised
completely by a set of classical probabilities.
Thus, entangled states bear correlations that do not exist in any
classical system.

\subsubsection{Multipartite entanglement}
\label{multient}

The definition of entangled states can straightforwardly be
generalised to {\it multipartite systems}, \ie systems that decompose
into more than two subsystems.
A $\nss$-partite system is described by a Hilbert space ${\cal H}$ that
decomposes into the tensor product of $p$ factor
spaces ${\cal H}_1\otimes\cdots\otimes{\cal H}_\nss$.
A pure state is separable if it can be written as a product of
$\nss$ states, each of which describes one of the subsystems -
any state that is not separable is entangled.
A mixed state is separable if it can be written as a convex sum of
product states, \ie of products of states each acting on a
single subsystem.
If it cannot, it is entangled.

For multipartite systems it is furthermore meaningful to distinguish between
different degrees of separability:
a pure state $\ket{\Psi}$ is called $k${\em -separable} if it can be written
as a product of $k$ states $\ket{\phi_i}$, each of which is an
element of one of the factor spaces, or of the product of some of those.
Thus a $k$-separable state with $k<\nss$ can contain entanglement
between some of the subsystems,
whilst there also are subsystems that are completely uncorrelated.
In this terminology, $\nss$-separability is equivalent to complete
separability.

\subsection{Separability criteria\label{secsepara}}
\subsubsection{Pure state separability - The Schmidt decomposition}
\label{secschmidt}

Pure bipartite states can be classified with the help of their
{\it Schmidt decomposition}.
Each bipartite pure state $\ket{\Psi}$ can be expressed in some product basis,
\be
\ket{\Psi}=
\sum_{ij}\ b_{ij}\ \ket{\varphi_i}\otimes\ket{\phi_j}\ .
\label{productbasis}
\ee
The local bases $\{\ket{\varphi_i}\}$ and $\{\ket{\phi_i}\}$ can be chosen
arbitrarily.
However, referring to a given state, there is always one distinguished basis.
It can be constructed with the following representations of the identity
operators:
${\mathbbm 1}_1=
\sum_i{\cal U}^\dagger\ket{\varphi_i}\bra{\varphi_i}{\cal U}$ acting
on the first factor space ${\cal H}_1$,
and, analogously ${\mathbbm 1}_2=
\sum_i{\cal V}^\dagger\ket{\phi_i}\bra{\phi_i}{\cal V}$,
acting on the second factor space ${\cal H}_2$.
${\cal U}$ and ${\cal V}$ are some arbitrary, local unitary
transformations on ${\cal H}_1$ and ${\cal H}_2$, respectively.
Inserting these identities in \Eq{productbasis},
the state $\ket{\Psi}$ can be expressed as
\be
\ket{\Psi}=\sum_{ij}\ [ubv]_{ij}\
\hspace{.1cm}{\cal U}^\dagger\ket{\varphi_i}
\otimes{\cal V}^\dagger\ket{\phi_j}\ ,
\ee
where the unitary matrices $u$ and $v$ are defined as
\be
u_{ij}=\matel{\varphi_i}{{\cal U}}{\varphi_j}
\hspace{.5cm}\mbox{and}\hspace{.5cm}
v_{ij}=\matel{\phi_j}{{\cal V}}{\phi_i}\ .
\footnote{Note that the order of indices in the definition of $v$ is
different from that of $u$.}
\ee
Now one can use the fact that every complex matrix $b$ can be diagonalised
by two unitary transformations $u$ and $v$, such that $ubv$, with real and non-negative diagonal elements $\sv_i$, provides the {\it singular value decomposition} of $b$~\cite{horn85}.
Hence, any pure state can be represented in terms of its
{\it Schmidt coefficients} $\sc_i=\sv_i^2$,
and of the associated {\it Schmidt basis}
$\ket{\sch_i}_1\otimes\ket{\sch_i}_2=
{\cal U}^\dagger\ket{\varphi_i}\otimes
{\cal V}^\dagger\ket{\phi_i}$:
\be
\ket{\Psi}=\sum_i\ \sqrt{\sc_i}\
\ket{\sch_i}_1\otimes\ket{\sch_i}_2\ ,
\label{schmidt}
\ee
with the sum limited by $d$, the dimension of the smaller subsystem.
Given that the Schmidt basis comprises - by construction - only separable
states,
all information about the entanglement of $\ket{\Psi}$ is now contained in the
Schmidt coefficients $\sc_i$.
The characterisation of all correlations of a given pure state is
therefore tantamount to the knowledge of all Schmidt coefficients.
The normalisation condition $\langle\Psi|\Psi\rangle=1$ implies that there are
$d-1$ independent coefficients.

The Schmidt coefficients can be easily computed with the help of one of the
{\it reduced density matrices} $\varrho_1=\tr_2\ket{\Psi}\bra{\Psi}$,
$\varrho_2=\tr_1\ket{\Psi}\bra{\Psi}$.
Assume, without loss of generality, $\dss=\dim({\cal
H}_1)\le\dim({\cal H}_2)$. Using \Eq{schmidt}, one easily verifies that the spectrum of
$\varrho_1$ is just given by the Schmidt coefficients -
the spectrum of $\varrho_2$ is given by the Schmidt coefficients and
$\dim({\cal H}_2)-\dim({\cal H}_1)$ vanishing eigenvalues.

The Schmidt coefficients also allow to distinguish separable from
entangled states -
a separable state is characterised by a vector of Schmidt coefficients
with only one non-vanishing entry:
$\vec\sc=\vec\sc_s=[1,0,\hdots,0]$,
whereas the {\it Schmidt vector} of an entangled state has at least
two non-vanishing components.
A state is called {\it maximally entangled}, if its Schmidt vector
reads $\vec\sc=\vec\sc_\me=[1/\dss,\hdots,1/\dss]$.
In Section~\ref{secmonotones} we will discuss in which respect this
terminology is legitimate.

It follows that the concept of Schmidt coefficients allows to relate the
degree of entanglement of pure bipartite states to the degree of mixing of the
corresponding reduced density matrices -
a pure reduced density matrix corresponds to a separable state,
whereas a maximally entangled state leads to a maximally mixed reduced
density matrix.

\subsubsection{Mixed state separability}
\label{mixedsep}

Whereas we have just seen that the separability of pure bipartite states can be
easily checked,
it turns out to be much more difficult to decide whether a given mixed
state bears quantum entanglement.
The above definition, \Eq{mixed-entangled-states}, for entangled mixed states
is not
constructive, and
generically it is not clear whether there is a set of product states
such that $\varrho$ can be represented as a convex sum of its elements.

The standard approach to decide on the separability of a given mixed
state relies on {\it positive maps}.
A map $\Lambda: {\cal B}({\cal H})\to {\cal B}({\cal H})$, where ${\cal
B}({\cal H})$ is the space of bounded linear operators on ${\cal H}$, is called {\it positive} if it maps positive operators on positive
ones, \ie
\be
\Lambda(\varrho)\ge 0,\hspace{.5cm}
\mbox{for all}\hspace{.5cm}\varrho\ge 0\ ,
\ee
where positivity of an operator $\varrho$ is just a short hand notation
stating that $\varrho$ is positive semi-definite,
\ie it has only non-negative eigenvalues.
A crucial property of positive maps is that a trivial extension
$\Lambda\otimes{\mathbbm 1}$ is {\em not} necessarily positive
\cite{Kra83}.
Consider a positive map
$\Lambda: {\cal B}({\cal H}_1)\mapsto {\cal B}({\cal H}_1)$:
if the trivial extension
$\Lambda\otimes{\mathbbm 1}$, with ${\mathbbm 1}$ the identity map
on ${\cal B}({\cal H}_2)$, is not positive,
it can be used to conclude on the separability of a mixed state
$\varrho$, acting on ${\cal H}_1\otimes{\cal H}_2$:
Since the extended map $\Lambda\otimes{\mathbbm 1}$
is not positive, there are some states $\eta$ such
that $\bigl(\Lambda\otimes{\mathbbm 1}\bigr)(\eta)\ngeq 0$.
However, if one assumes the considered state $\varrho$ to be separable,
its convex decomposition into product states (\ref{mixed-separable-states})
implies
\be
\bigl(\Lambda\otimes{\mathbbm 1}\bigr)(\varrho)=
\sum_i\ p_i\ \Lambda(\varrho_1^i)\otimes\varrho_2^i\ .
\ee
Obviously, any expectation value of this quantity is non-negative,
such that $\bigl(\Lambda\otimes{\mathbbm 1}\bigr)(\varrho)\ge 0$.
Equivalently, a state $\varrho$ is entangled if
$\bigl(\Lambda\otimes{\mathbbm 1}\bigr)(\varrho)\ngeq 0$.

However, the inverse statement does not hold in general.
The mere fact that $\varrho$ remains positive under the extended map
does not necessarily imply that $\varrho$ is separable.
Only if
\be
\bigl(\Lambda\otimes{\mathbbm 1}\bigr)(\varrho)\ge 0
\hspace{.5cm}\mbox{{\it for all} positive maps }\Lambda\ ,
\ee
one may conclude that $\varrho$ is separable~\cite{hor96}.
Note that for the complementary implication alone one only needs to
find one positive map $\Lambda$ with
$\bigl(\Lambda\otimes{\mathbbm 1}\bigr)(\varrho)\ngeq 0$.
This statement does not allow to derive a sufficient separability criterion for
the very general case,
since the classification of positive maps is still an unsolved
problem.
A large\footnote{`large' in the topological sense: the set of states
with negative partial transpose contains an open
set.} class of entangled states is detected by the special choice of
the transposition $T=:\Lambda$~\cite{peres} - that indeed is a positive map.
The {\it partial transpose}
$\varrho^{pt}=\bigl(T\otimes{\mathbbm 1}\bigr)(\varrho)$
of a state $\varrho$ is deduced as the relevant auxiliary quantity:
if $\varrho^{pt}$  has at least one negative eigenvalue,
\be
\varrho^{pt}\ngeq 0\ ,
\label{ppt}
\ee
the state $\varrho$ is entangled.

However, if $\varrho^{pt}$ is positive, one can infer separability
of $\varrho$ only for low-dimensional, namely $2\times 2$ and $2\times 3$
systems. For these, the positive partial transpose ({\it ppt})
or {\em standard criterion} unambiguously
distinguishes separable and entangled states~\cite{hor96}.
However, in higher dimensions there exist entangled
states~\cite{hor98,hor99} that are not
detected by the ppt criterion.

\subsection{How to quantify entanglement? \label{howtoquant}}
Since the definition of entangled states given in \Eq{mixed-entangled-states} is not
constructive, it turned out difficult to decide whether a given state
is separable - and the general solution of this problem is still unknown.
Moreover, the non-constructive definition also complicates finding a
{\em quantitative} description of entanglement, rather than the purely
qualitative one.
How can you measure something, if you don't even know what it is?

The basic idea for a quantitative treatment is to classify all kinds
of operations that in principle can be applied to quantum systems
and that can create or increase only classical correlations, but none
of quantum nature.
Any quantity that is supposed to quantify entanglement needs to be
monotonously decreasing under such operations~\cite{vidal00,ben96b}.

In our subsequent discussion,
we will not distinguish between operations describing
the time evolution of a real system,
and those which serve just as a mathematical tool.
In the latter case one can always have in mind a Gedankenexperiment
where the considered
operation is implemented.
For the following considerations it is not crucial whether one has the
technical prerequisites and experimental skills to perform a
considered operation -
but rather that the operation is in principle allowed
by the laws of quantum mechanics.
Therefore, we do not consider technical problems - as long as we do not
refer to real experiments.

A map
${\cal E}:\ {\cal B}({\cal H}_i)\to{\cal B}({\cal H}_f)$
describing the evolution of a quantum system has to be linear,
\be
{\cal E}(\lambda \varrho+\lambda^{\prime}\varrho^{\prime})=
\lambda\ {\cal E}(\varrho)+\lambda^{\prime}\ {\cal E}(\varrho^{\prime})\ ,
\ee
due to the underlying linear Schr\"odinger equation.
Moreover, in order to ensure positivity of $\varrho$,
any map ${\cal E}$ has to be positive.
However, this requirement is not strong
enough to ensure positivity of $\varrho$ in all cases.
Since one can always consider a system as a subsystem of a larger
one, one has to allow for extensions
${\cal E}\otimes{\mathbbm 1}$ of ${\cal E}$.
The extended map acts on the entire system in such a way that the
original map affects the considered subsystem,
whereas the identity map acts on the residual system degrees of
freedom.
As already mentioned in Section~\ref{mixedsep},
such a trivial extension does {\it not} necessarily preserve positivity.
In order to guarantee positivity of the entire system state, one has to
require that the described extension be a positive map for identity
maps ${\mathbbm 1}$ in any dimension,
\ie that ${\cal E}$ is {\it completely positive}.
Consequently, any evolution consistent with the general rules of quantum
mechanics can be described by a linear, completely positive map, called
{\it quantum operation}.

A unitary evolution is just a special case of such a quantum operation
- general quantum operations can also describe non-unitary
evolutions, {\it e.g.} due to environment coupling or measurements.
Any such quantum operation can be composed from elementary
operations~\cite{Kra83,Kra71,nielsen00}:
\begin{itemize}
\item[-]unitary transformations,
${\cal E}_1(\varrho)=U\varrho\ U^\dagger$\ ;
\item[-]addition of an auxiliary system,
${\cal E}_2(\varrho)=\varrho\otimes\sigma$, where $\varrho$ is the
original system and $\sigma$ is the auxiliary one;
\item[-]partial traces,
${\cal E}_3(\varrho)=\tr_p\varrho$\ ;
\item[-]projective measurements,
${\cal E}_4(\varrho)={P_k\varrho P_k}/\tr(P_k\varrho)$\ , with $P_k^2=P_k$,
\end{itemize}
which allows for a physical interpretation thereof.
For a formal mathematical treatment it is useful to note
that any quantum operation can be expressed as an operator sum~\cite{hk69,hk70}
\be
{\cal E}(\varrho)=\sum_i\ E_i\varrho E_i^\dagger\ ,
\hspace{.5cm}\mbox{with}\hspace{.5cm}
\sum_iE_i^\dagger E_i={\mathbbm 1}\ ,
\label{opsum}
\ee
with suitably defined linear operators $E_i$.

The reduced dynamics of a system initially prepared in the state
$\varrho(0)$, coupled to an environment with initial state $\ket{e_0}$
can be interpreted in terms
of quantum operations.
If we allow for an interaction between system and environment,
we will have a unitary evolution ${\cal U}(t)$ in both system and
environment.
The system state after time $t$ is obtained by evolving the
system-bath state over $t$, followed by a trace over the environment:
\be
\varrho(t)=\tr_{\mbox{\rm{env}}}\Bigl(\
{\cal U}(t)\ \ket{\e_0}\bra{\e_0}\otimes
\varrho(0)\ {\cal U}^\dagger(t)\ \Bigr)\ .
\ee
Expressing the trace over the environmental degrees of freedom by a
sum over an orthonormal basis $\{\ket{\chi_i}\}$,
one immediately obtains the above operator sum representation
\be
\varrho(t)=\sum_i\ E_i(t)\ \varrho(0)\ E_i^\dagger(t)\ ,
\hspace{.5cm}\mbox{with}\hspace{.5cm}
E_i=\matel{\chi_i}{{\cal U}(t)}{\e_0}\ ,
\ee
where the operators $E_i(t)$ satisfy the resolution of the identity required
in \Eq{opsum}.

For our purposes, it will be useful to distinguish the
following types of quantum operations:
\begin{itemize}
\item[-]local operations,
\item[-]global operations,
\item[-]local operations and classical communication (LOCC).
\end{itemize}

\subsubsection{Local operations}

An operation is called {\it local} if under its action the subsystems
evolve independently from each other.
In terms of operator sums this is expressed (for bipartite systems,
with a straightforward generalisation for the multipartite case) as
\be
{\cal E}_{loc}(\varrho)=\sum_{ij}
E_i\otimes F_j\ \varrho\ E_i^\dagger\otimes F_j^\dagger,
\hspace{.3cm}\mbox{with}\hspace{.3cm}
\sum_{i,j}E_i^\dagger E_i\otimes F_j^\dagger F_j=
{\mathbbm 1}_{{\cal H}_1\otimes{\cal H}_2}\ .
\label{Eloc}
\ee
Local unitary evolutions
${\cal U}_{loc}={\cal U}_1\otimes{\cal U}_2$
are just special cases of general local operations.
Since both subsystems evolve independently from each other,
possibly preexisting correlations remain unaffected.
A product state will remain a product state,
\be
{\cal E}_{loc}(\varrho_1\otimes\varrho_2)=
\Bigl(\sum_iE_i\varrho_1E_i^\dagger\Bigr)\otimes
\Bigl(\sum_iF_i\varrho_2F_i^\dagger\Bigr),
\ee
and any separable state will remain separable under local operations:
\be
{\cal E}_{loc}\Bigl(\sum_i\ p_i\ \varrho_1^i\otimes\varrho_2^i\Bigr)=
\sum_i\ p_i\
\Bigl(\sum_iE_i\varrho_1^iE_i^\dagger\Bigr)\otimes
\Bigl(\sum_iF_i\varrho_2^iF_i^\dagger\Bigr).
\ee
Therefore, starting from a separable state no correlations -
neither classical nor quantum -
can be created by local operations alone.

\subsubsection{Global operations}

If two subsystems are interacting with each other, their evolution will
in general not derive from purely local operations.
Any operation that is not local is called {\it global}.
Under this type of operations all kinds of correlations can increase, as
well as decrease.
Therefore, entangled states can be created from initially separable
states and vice versa.
The most prominent and natural way of creating entangled states is a
global unitary evolution due to an interaction between subsystems.

\subsubsection{Local operations and classical communication (LOCC)}
\label{seclocc}

A prominent subclass of global operations are
{\it local operations and classical communication} (LOCC).
They comprise general local operations, and also allow
for classical correlations between them.
The idea behind it is to allow arbitrary local operations and, in
addition, to admit all classical means to correlate their application.
Hence, parties having access to different subsystems can use means of
classical communication to exchange information about their locally performed
operations and the respective outcomes, and, subsequently, apply some
further local operations {\it conditioned} on the communicated information.

In terms of operator sums this can be expressed as\footnote{
Strictly speaking, \Eq{locc} characterises {\em separable} operations
that {\em include} LOCC-operations.
Though not every separable operation is LOCC.
The exact definition of LOCC reads
${\cal E}_{\mbox{\tiny LOCC}}(\varrho)=\sum_{ijk\hdots}\hdots 
(\un\otimes F_{ijk})
(E_{ij}\otimes\un)
(\un\otimes F_i)\varrho(\un\otimes F_i^\dagger)
(E_{ij}^\dagger\otimes\un)
(F_{ijk}^\dagger\otimes\un)\hdots$\cite{wot99}}
\be
{\cal E}(\varrho)=\sum_i
E_i\otimes F_i\ \varrho\ E_i^\dagger\otimes F_i^\dagger,
\hspace{.3cm}\mbox{with}\hspace{.3cm}
\sum_iE_i^\dagger E_i\otimes F_i^\dagger F_i={\mathbbm 1}_{{\cal H}_1\otimes{\cal H}_2}\ .
\label{locc}
\ee
In contrast to \Eq{Eloc}, only a single sum is involved in the
description of LOCC operations.
This is a manifestation of the correlated application of the
respective operations on the subsystems:
if the operator $E_i$ is applied to the first subsystem,
the operator $F_i$ is applied to the second subsystem.

LOCC operations can be used to create classical correlations
between subsystems.
In general, a product state will not remain a direct product under the
action of an LOCC operation:
\be
{\cal E}_{\mbox{\tiny LOCC}}(\varrho_1\otimes\varrho_2)=
\sum_i\left(E_i\varrho_1E_i^\dagger\right)\otimes
\left(F_i\varrho_2F_i^\dagger\right)=
\sum_i\ p_i\ \varrho_1^i\otimes\varrho_2^i\ ,
\ee
with
$\varrho_1^i=E_i\varrho_1E_i^\dagger/\tr\left(E_i\varrho_1E_i^\dagger\right)$,
$\varrho_2^i=F_i\varrho_2F_i^\dagger/\tr\left(F_i\varrho_2F_i^\dagger\right)$,
and $p_i=\tr\left(E_i\varrho_1E_i^\dagger\right)\ \tr\left(F_i\varrho_2F_i^\dagger\right)$.
Thus, classical probabilistic correlations can change under the action of
LOCC operations.
However, any separable state will always remain separable under LOCC
operations.
Accordingly, entangled states cannot be created with LOCC operations.

\subsection{Entanglement monotones\label{secmonotones}}
Since we have argued that entanglement cannot be created using LOCC
operations, our discussion at the beginning of
Section~\ref{howtoquant} suggests to consider quantities that do not
increase precisely under LOCC operations to quantify entanglement. Any scalar valued function that satisfies this criterion is called an
{\it entanglement mo\-no\-tone}~\cite{vidal00,ben96b}.

\subsubsection{Pure bipartite states\label{purbip}}

For pure bipartite states there exists a simple criterion that allows for the
characterisation of entanglement monotones.
It was shown that a state $\ket{\Xi}$ can be prepared
starting from a second state $\ket{\Phi}$ and using only LOCC,
if and only if the vector
$\vec\sc_\Xi$ of Schmidt coefficients of $\ket{\Xi}$ majorises
$\vec\sc_\Phi$~\cite{Ni99} 
\be
\vec\sc_\Phi\prec\vec\sc_\Xi\ .
\ee
Majorisation means that the components
$[\sc_\Xi]_i$ and $[\sc_\Phi]_i$
of both vectors, listed in nonincreasing order, satisfy
$\sum_{i=1}^j[\sc_\Xi]_i\geq\sum_{i=1}^j[\sc_\Phi]_i$, for $1<j\le
\dss$, with equality when $j=\dss$ (due to normalisation, see
Eq.~(\ref{schmidt})). 
Since the Schmidt vector $\vec\lambda_\me$, with equal components
$1/\dss$ as introduced in Section~\ref{secschmidt}, is majorised by any
vector $\vec\lambda$, any bipartite state can be prepared with LOCC
starting out from a state $\ket{\Psi_\me}$ with Schmidt vector
$\vec\lambda_\me$.
This justifies calling $\ket{\Psi_\me}$ `maximally entangled'.

Since entanglement cannot increase under LOCC operations,
any mo\-no\-tone ${\cal M}$ has to satisfy
\be
{\cal M}(\Phi)\ > \ {\cal M}(\Xi)\ ,
\hspace{.5cm}\mbox{for}\hspace{.5cm}
\vec\sc_\Phi\prec\vec\sc_\Xi\ .
\ee 
This condition is known as {\it Schur concavity}.
It is satisfied if and only if~\cite{An89}
${\cal M}$ given as a function of the Schmidt coefficients is
invariant under permutations of any two arguments and satisfies
\be
(\sc_1-\sc_2)
\left(
\ds\frac{\partial{\cal M}}{\partial\sc_1}-
\ds\frac{\partial{\cal M}}{\partial\sc_2}\right)\le 0\ .
\label{schur}
\ee
Due to the above-mentioned invariance, there is nothing peculiar about
the first two components of $\vec\lambda$ -
if \Eq{schur} holds true for $\lambda_1$ and $\lambda_2$,
it is satisfied for any two components of $\vec\lambda$. 

The above characterisation allows to derive several entanglement mo\-no\-tones for pure
states.
Very useful quantities in this context are the reduced density
matrices, $\varrho_1$ or $\varrho_2$, obtained by tracing over one
subsystem
\be
\varrho_1=\tr_2\ket{\Psi}\bra{\Psi}\ , \hspace{1cm}
\varrho_2=\tr_1\ket{\Psi}\bra{\Psi}\ .
\ee
The basic idea is that the degree of mixing of a reduced density
matrix is directly related to the amount of entanglement of the pure
state $\ket{\Psi}$.
Any function $g(\varrho_r)$ of a reduced density matrix $\varrho_r$
that is
\begin{itemize}
\item[-]invariant under unitary transformations,
$g(\varrho_r)=g({\cal U}\varrho_r{\cal U}^\dagger)$, and
\item[-]concave,
$g(\varrho_r)\ge\lambda g(\varrho_\alpha)+(1-\lambda)g(\varrho_\beta)$,
for any $0\le\lambda\le 1$, and states
$\varrho_\alpha$ and $\varrho_\beta$ such that
$\varrho_r=\lambda\varrho_\alpha+(1-\lambda)\varrho_\beta$,
\end{itemize}
is Schur concave, and therefore provides an entanglement monotone
${\cal M}(\Psi)=g(\varrho_r)$~\cite{vidal00}.
The most prominent choice of $g$ is the von Neumann entropy
\be
S(\varrho_r)=-\tr\varrho_r\ln\varrho_r
\label{vNentr}
\ee
of the reduced density matrix,
often simply called {\it the} entanglement $E(\Psi)=S(\varrho_r)$ of
the pure state $\ket{\Psi}$.

Note that due to the invariance of $g$ under unitary transformations,
$g$ can only be a function of unitary invariants, hence,
of the spectrum of $\varrho_r$.
Accordingly, it is not necessary to distinguish between $\varrho_1$ and $\varrho_2$,
since they have the same non-vanishing eigenvalues.
If both subsystems have the same dimensions, the spectrum of
$\varrho_1$ equals that of $\varrho_2$.
If the dimensions are not equal the reduced density matrix of the
larger subsystem has some additional vanishing eigenvalues.
That is why one often does not distinguish between $\varrho_1$ and
$\varrho_2$,
but rather expresses ${\cal M}(\Psi)$ as
\be
{\cal M}(\Psi)=g(\varrho_r)\ ,
\hspace{.5cm}\mbox{with}\hspace{.5cm}
\varrho_r=\tr_p\ket{\Psi}\bra{\Psi}\ ,
\label{subsys}
\ee
`the' reduced density matrix, where the partial trace $\tr_p$
does not specify explicitly which subsystem is traced out.
In subsection~\ref{symmetries} we will discuss a situation where the proper
choice of the subsystem over which the trace is performed is not completely arbitrary.

Finally, it should be kept in mind that 
a single monotone is in general insufficient to completely
characterise the quantum correlations contained in a given pure
state.
For such a characterisation, knowledge of {\it all} -
{\it i.e.} $\dss-1$
(with $\dss = \min( \dim({\cal H}_1), \dim({\cal H}_2))$)
independent Schmidt coefficients is required.
For pure states this does not represent a serious problem but already
indicates that, with increasing dimension, the complete
characterisation of arbitrary mixed states will define a task of
rapidly increasing complexity.
Further down in this review, when dealing with higher dimensional
mixed states, we will therefore have to specify which specific type of
correlations we want to scrutinize, rather than to give an  
exhaustive description of all
correlations inscribed into a given state.

\subsubsection{Mixed states}
\label{mixedstates}

For pure bipartite states it is rather simple to find some entanglement
monotones - any unitarily invariant, concave function of the reduced
density matrix defines one.
This is due to the fact that there are no classical probabilistic
correlations contained in pure states.
For mixed states the situation is much more involved, because there
are both classical and quantum correlations that have to be
discriminated against each other by an entanglement monotone.
It is by no means obvious to devise a unique generalisation of a pure state
monotone ${\cal M}(\Psi)$ to a mixed state monotone ${\cal M}(\varrho)$,
such that
\begin{itemize}
\item[-] ${\cal M}(\varrho)$ reduces to the original pure state
  definition when applied to pure states, and
\item[-] ${\cal M}(\varrho)$ is an entanglement monotone, \ie
non-increasing under LOCC.
\end{itemize}

We will here follow one particular generalisation that applies to any pure
state monotone~\cite{vidal00,ben96b}, and therefore is the most
commonly used one.
It can be easily formulated, but poses severe problems when it comes
to its quantitative evaluation.
Any mixed state can be expressed as a convex sum of pure states:
\be
\varrho=\sum_i\ p_i\ \ket{\Psi_i}\bra{\Psi_i}\ .
\label{mixdec}
\ee
On a first glance, it might appear as a self-evident generalisation to sum up
the entanglement assigned by a certain monotone to the pure states in
\Eq{mixdec},
weighted by the prefactors $p_i$.
Unfortunately, the decomposition into pure states is not unique,
and different decompositions in general lead to different values for a
given entanglement monotone.
The proper, unambiguous generalisation of a pure state monotone,
that we will also use in the following, therefore uses the
infimum over all decompositions into pure states - the so-called
{\it convex roof}~\cite{uhl00}
\be
{\cal M}(\varrho)=\inf_{\{p_i,\Psi_i\}}\sum_i\ p_i\ {\cal M}(\Psi_i)\ ,
\hspace{.2cm}\mbox{with}\hspace{.2cm}p_i>0\ ,\hspace{.2cm}
\mbox{and}\hspace{.2cm}
\varrho=\sum_ip_i\ket{\Psi_i}\bra{\Psi_i}\ .
\label{roof}
\ee
An explicit evaluation of this quantity for a specific state -
one has to find the infimum over all possible decompositions into pure
states -
implies a high dimensional optimisation problem - in general
a very hard computational task.

To ease this enterprise, it is convenient to
make use of the following characterisation of all ensembles of pure
states which represent a certain mixed state.
Using {\it subnormalised states}
\be
\ket{\psi_i}=\sqrt{p_i}\ \ket{\Psi_i}
\label{subnormalised}
\ee
allows to reduce the number of involved quantities.
Since the $p_i$ are positive, one has
$\ket{\psi_i}\bra{\psi_i}=p_i\ket{\Psi_i}\bra{\Psi_i}$.
Assume one ensemble $\{\ket{\psi_i}\}$ is known such that
$\varrho=\sum_i\ket{\psi_i}\bra{\psi_i}$ - {\it e.g.} the eigensystem of
$\varrho$ .
New ensembles defined as
\be
\ket{\phi_i}=\sum_jV_{ij}\ket{\psi_j},
\hspace{.5cm}\mbox{with}\hspace{.5cm}
\sum_iV_{ki}^\dagger V_{ij}=\delta_{jk}\ ,
\label{ensemble}
\ee
represent the same mixed state
$\varrho=
\sum_i\ket{\psi_i}\bra{\psi_i}=
\sum_i\ket{\phi_i}\bra{\phi_i}$,
and any ensemble representing $\varrho$ can be constructed in this way~\cite{wot93a,schroed}.
In the following, any matrix $V$ satisfying \Eq{ensemble} will be
referred to as {\it left unitary}.
The number, {\it cardinality}, of ensemble members of the decomposition of
$\varrho$ (i.e., the length of the index set of $i$ in Eq.~(\ref{ensemble}))
is not fixed by the rank, i.e., the number of nonvanishing eigenvalues of the
density matrix which represents the considered state.
There is no {\it a priori} maximum cardinality,
though it is sufficient to consider ensembles
with cardinality not larger than the square of the considered
state's rank~\cite{uhl98}.
However, there is no evidence that it is necessary to employ ensembles
of this maximum cardinality, and, in particular, 
there is no proof that the infimum in \Eq{roof} cannot be found with smaller 
ensembles.
Nonetheless, without a sharper bound on the length of the
decomposition, we need to find the optimal left-unitary matrix
$V\in{\mathbbm C}^{n^2\times n}$, which
implies an optimisation procedure of dimension $\sim n^3$ to compute the
entanglement ${\cal M}(\varrho)$ of a given state $\varrho$ of rank $n$.
Since there is no simple parametrisation of arbitrary left-unitary
matrices,
the constraint
$\sum_iV_{ki}^\dagger V_{ij}=\delta_{jk}$ even complicates numerical
implementation.

\subsection{Entanglement measures\label{secmeasures}}
Entanglement monotones that satisfy some additional axioms
are called {\em entanglement measures} $E$.
So far, however, there is no uniquely accepted list of axioms,
hence there is no commonly accepted distinction between monotones
and measures~\cite{ple98,bru02}.
We do not attribute too much relevance to this question of terminology,
and just present here a list of potential axioms:

\begin{itemize}
\item[-]$E(\varrho)$ vanishes exactly for separable states.
\item[-]{\it additivity}: the entanglement of several copies of a state
  adds up to $n$ times the entanglement of a single copy,
  $E(\varrho^{\otimes n})=nE(\varrho)$.
\item[-]{\it subadditivity}: the entanglement of two states is not
  larger than the sum of the entanglement of both individual states,
  $E(\varrho\otimes\varrho^{\prime})\le E(\varrho)+E(\varrho^{\prime})$.
\item[-]{\it convexity}:
  $E(\lambda\varrho+(1-\lambda)\varrho^{\prime})\le\lambda
  E(\varrho)+(1-\lambda)E(\varrho^{\prime})$, for $0\le\lambda\le 1$.
\end{itemize}
Some authors additionally require that an entanglement
measure has to be invariant under local unitary transformations.
However, this is already implied by monotonicity under LOCC (Section~\ref{secmonotones}),
since monotonicity implies invariance under transformations that are
invertible within the class of LOCC operations~\cite{Ho02}.
Local unitaries and their inverses are LOCC operations.
Thus, any entanglement monotone and measure has to be non-increasing
under both the former and the latter.
Since non-increasing behaviour under the latter implies non-decreasing
behaviour under the former and vice versa,
any monotone or measure has to be invariant under local unitaries.

There are attempts to find a distinct set of axioms that leads to a
unique measure~\cite{hor03}.
On the other hand, it is sometimes necessary to relax some of the
above listed constraints, in order to find a measure that is computable.
For example, {\it negativity}~\cite{vidal02a}
has become a commonly used quantity
although it vanishes for a class of entangled states~\cite{vidal02a}.
Though, compared to other measures it has the major advantage
that it can be computed straightforwardly.

\section{Concurrence\label{secwotconc}}

\subsection{Two-level systems\label{concurrence-two-level}}
{\it Concurrence} was originally introduced as an auxiliary quantity,
used to calculate the {\it entanglement of formation} of
$2\times 2$ systems.
However, concurrence can also be considered as an independent
entanglement measure~\cite{wot98}.
The original definition of concurrence~\cite{ben96b,wot97} for
bipartite two-level systems is given in terms of a special basis
\be
\ket{e_1}=\ket{\Phi^+}\ ,\hspace{.5cm}
\ket{e_2}=i\ket{\Phi^-}\ ,\hspace{.5cm}
\ket{e_3}=i\ket{\Psi^+}\ ,\hspace{.5cm}
\ket{e_4}=\ket{\Psi^-}\ ,
\ee
where
$\ket{\Phi^{\pm}}=\left(\ket{00}\pm\ket{11}\right)/\sqrt 2$
and
$\ket{\Psi^{\pm}}=\left(\ket{01}\pm\ket{10}\right)/\sqrt 2$
are the {\it Bell states}~\cite{ben96b}.
Using this particular basis, the concurrence $c$ of a pure state
$\ket{\Psi}$ is defined as
\be
c(\Psi)=\Bigl|\sum_i\ovl{e_i}{\Psi}^2\Bigr|\ .
\label{wotaltcon}
\ee
Writing this definition more explicitly,
$c(\Psi)=\bigl|\sum_i
\ovl{\Psi^\ast}{e_i^\ast}\ovl{e_i}{\Psi}\bigr|$,
one ends up, after summation, with the alternative formulation~\cite{wot98}
\be
c(\Psi)=\Bigl|\matel{\Psi^\ast}{\sigma_y\otimes\sigma_y}{\Psi}\Bigr|\ ,
\label{wotcon}
\ee
where $\sigma_y$ is the second Pauli matrix,
and $\ket{\Psi^\ast}$ is the complex conjugate of $\ket{\Psi}$ with the
conjugation performed in the standard (real) basis
$\{\ket{00},\ \ket{01},\ \ket{10},\ \ket{11}\}$.
Since a scalar product implies a complex conjugation anyway,
the second conjugation cancels the first one
such that
$\bra{\Psi^\ast}=\sum_{ij}\Psi_{ij}\bra{ij}$ is the transpose and
not the adjoint of
$\ket{\Psi}=\sum_{ij}\Psi_{ij}\ket{ij}$.
\Eq{wotcon} is the most commonly used formulation and is often considered as
the definition of concurrence rather than \Eq{wotaltcon}.

The concurrence of mixed states is given by the corresponding convex
roof, alike \Eq{roof}:
\be
c(\varrho)=\inf_{\{p_i,\Psi_i\}}\sum_i\ p_i\ c(\Psi_i)\ ,
\hspace{.2cm}\mbox{with}\hspace{.2cm}p_i>0\ ,\hspace{.2cm}
\mbox{and}\hspace{.2cm}
\varrho=\sum_ip_i\ket{\Psi_i}\bra{\Psi_i}\ .
\label{mixwotcon}
\ee
Concurrence of a pure subnormalised state $\ket{\psi}$ (see
Eq.~(\ref{subnormalised})) can be expressed as
$c(\psi)=|f_c(\psi,\psi)|$, in terms of the function
\be
f_c(\psi_j,\psi_k)=
\matel{\psi_k^\ast}{\sigma_y\otimes\sigma_y}{\psi_j}\ ,
\ee
that is linear in both arguments.
These linearity properties, together with the parametrisation of all decompositions of
$\varrho$ into pure states given in \Eq{ensemble}, allows to
write \Eq{mixwotcon} as
\be
c(\varrho)=\inf_{V}\sum_{i}
\Bigl|\sum_{j,k}V_{ij}f_c(\psi_j,\psi_k)
\left[V^T\right]_{ki}\Bigr|\ .
\ee
The quantities $f_c(\psi_j,\psi_k)$
can be understood as elements $\tau_{jk}$
of a complex symmetric matrix $\tau$.
Hence one can use the compact matrix notation
\be
c(\varrho)=\inf_{V}\sum_i\Bigl|\left[V\tau V^T\right]_{ii}\Bigr|\ .
\label{wotinf}
\ee
The infimum of this quantity is known~\cite{wot98} to be given by
\be
c(\varrho)=\max\left\{\sv_1-\sum_{i=2}^4\ \sv_i,\ 0\right\}\ ,
\label{wotinfsol}
\ee
where the $\sv_i$ are the singular values of $\tau$, in decreasing
order.
They can be obtained as the square roots of the eigenvalues of the
positive hermitian matrix $\tau\tau^\dagger$.\\

Since we will refer to infima of expressions similar to that given in
\Eq{wotinf} several times later on,
we discuss the derivation of this infimum in some detail.
Though, we do not follow here the original derivation~\cite{wot98},
but rather present a generalisation~\cite{uhl00} valid in
arbitrary dimensions -
which we will need for later reference when considering subsystems
with more than two levels.
Thus, the following considerations do not only apply to the hitherto
discussed $2\times 2$ case, but also to systems of arbitrary finite
dimension.\\

Any complex matrix $M\in{\mathbbm C}^{n_1\times n_2}$ can be
diagonalised~\cite{horn85} as
\be
M=U_l D U_r\ ,
\ee
where $U_l\in{\mathbbm C}^{n_1\times \dss}$ and
$U_r\in{\mathbbm C}^{\dss\times n_2}$, with $\dss=\min(n_1,n_2)$,
are, respectively, left- and right unitary,
\ie $U_l^\dagger U_l={\mathbbm 1}$ and
$U_rU_r^\dagger={\mathbbm 1}$,
and $D\in{\mathbbm C}^{\dss\times \dss}$ is a diagonal matrix with real and
positive diagonal elements, referred to as singular values of $M$.
Moreover, $U_l$ and $U_r$ can always be chosen such that the
singular values are arranged in decreasing order along the diagonal.

Applying this to a square, complex symmetric matrix $\tau$,
one concludes that $\tau$ can be diagonalised with a unitary transformation $U$ as
\be
U\tau U^T=\mbox{diag}\bigl[\sv_1,\hdots,\sv_n\bigr]=:\tau^{\mbox{diag}}\ .
\ee
Given this diagonal representation,
one defines a transformation $V_H$
with the help of $n\times 2^k$ {\it Hadamard Matrices}
$H$~\cite{weis} with $2^k\ge n$. A square $2^k\times 2^k$ Hadamard matrix can be
constructed for each $k$ and, by definition, has its columns given by
mutually orthogonal real vectors, with the same absolute value
$1/(2^{k/2})$ of all elements. We shall denote by $H$ (and call it
also a Hadamard matrix) a rectangular $n\times 2^k$ matrix obtained
from the original square one by keeping only $n$ rows.
Due to the rows' orthogonality, $H$ is left unitary,
$H^\dagger H={\mathbbm 1}$.

The transformation matrix $V_H$ is a modification of a Hadamard
matrix.
Namely each column $j$ but the first one is multiplied
with a phase factor $ie^{i\varphi_j/2}$.
The latter does not affect left-unitarity,
$V_H^\dagger V_H={\mathbbm 1}$.
However, since $V_H^T$ enters \Eq{wotinf}
instead of $V_H^\dagger$, the phase factors are indeed important.
Carrying out the transformation, one obtains
\be
\left[V_H\tau^{\mbox{diag}}V_H^T\right]_{ii}=
\frac{1}{2^k}\bigl(
\sv_1-\sum_{j>1}\sv_je^{i\varphi_j}
\bigr)\ ,
\label{complexsing}
\ee
where one needs not care about the non-diagonal entries,
since only diagonal elements are summed up in the end.

So far we found a (left-unitary) transformation $V_HU$ such that
\be
\sum_i\Bigl|\left[V_HU\tau U^TV_H^T\right]_{ii}\Bigr|=
\Bigl|\sv_1-\sum_{j>1}\sv_je^{i\varphi_j}\Bigr|\ .
\ee
\begin{figure}
\psfrag{sep}{{\bf separable state}}
\psfrag{ent}{{\bf entangled state}}
\psfrag{R}{$\Re$}
\psfrag{I}{$\Im$}
\psfrag{l1}{$\sv_1$}
\psfrag{l2s}{$-\sv_2e^{i\varphi_2}$}
\psfrag{l3s}{$-\sv_3e^{i\varphi_3}$}
\psfrag{l4s}{$-\sv_4e^{i\varphi_4}$}
\psfrag{l5s}{$-\sv_5e^{i\varphi_5}$}
\psfrag{l2}{$-\sv_2$}
\psfrag{l3}{$-\sv_3$}
\psfrag{l4}{$-\sv_4$}
\psfrag{l5}{$-\sv_5$}
\epsfig{file=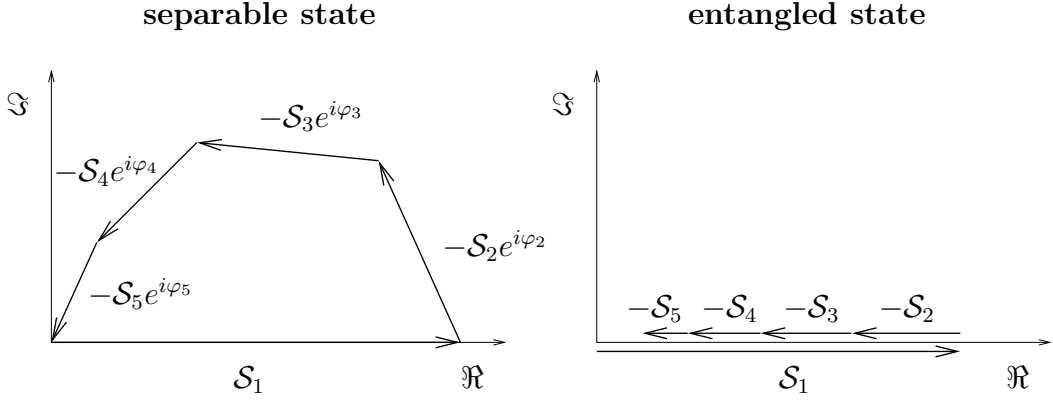,width=1.0\textwidth,angle=0}
\caption{
  Complexified singular values $\sv_1$ and $-\sv_je^{i\varphi_j}$
  ($j>1$) - see \Eq{complexsing} - plotted in the complex plane.
  For a separable state (left) one can always find appropriate phase factors such
  that all terms add up to $0$.
  However, for an entangled state (right) this is not possible.
  The minimum of \Eq{complexsing} can be obtained for
  $e^{i\varphi_j}=1$ for all $j$.
}
\label{kette}
\end{figure}
Now one has to distinguish two cases.
In the first case one has $\sv_1\le\sum_{j>1}\sv_j$.
In that case one can always find phases $\varphi_j$ such that
$\sv_1-\sum_{j>1}\sv_je^{i\varphi_j}=0$,
as depicted in Fig.~\ref{kette}.
In the second case, where $\sv_1>\sum_{j>1}\sv_j$,
the optimal choice for all phases (such that
$\sv_1-\sum_{j>1}\sv_je^{i\varphi_j}$ be minimal) is $\varphi_j=0$, and one gets
$\sv_1-\sum_{j>1}\sv_j$.
Altogether, we found a transformation such that
\be
\sum_i\Bigl|\left[V_HU\tau U^TV_H^T\right]_{ii}\Bigr|=
\max\left\{\sv_1-\sum_{j>1}\sv_j,0\right\}\ .
\label{infimum}
\ee
It is now easy to show that there is no left unitary transformation
leading to a smaller result.
We can restrict ourselves to the second case above,
$\sv_1>\sum_{j>1}\sv_j$,
since $0$ is the smallest possible value of a non-negative quantity anyway.
To do so, we start with the diagonal form $\tau^{\mbox{diag}}$ of $\tau$, and
transform it with
$U_p=\mbox{diag}(1,i,\hdots,i)$ such that we obtain
\be
\tilde\tau^{\mbox{diag}}=U_p\tau^{\mbox{diag}}U_p^T=\mbox{diag}\bigl[\sv_1,-\sv_2,\hdots,-\sv_n\bigr]\
.
\ee
Now one has
\be
\sum_i\Bigl|\sum_{jk}V_{ij}\tilde\tau^{\mbox{diag}}_{jk}V_{ki}^T\Bigr|\ge
\Bigl|\sum_i\bigl(V_{i1}^2\sv_1-
\sum_{j>1}V_{ij}^2\sv_j\bigr)\Bigr|\ .
\ee
One can always choose $V$ in such a way that the $V_{i1}$ are real for any $i$.
Then one has
$\sum_iV_{i1}^2=1$, and $\left|\sum_iV_{i1}^2\right|\le 1$.
Therefore:
\be
\sum_i\Bigl|\sum_{jk}V_{ij}\tilde\tau^{\mbox{diag}}_{jk}V_{ki}^T\Bigr|\ge
\sv_1-\sum_{i>1}\sv_i\ .
\ee
Thus we have shown that \Eq{wotinfsol} really is the infimum
corresponding to \Eq{wotinf}.

From this exact expression for the concurrence for mixed states, one can also
deduce the entanglement of formation.
As we have seen in Section~\ref{secmonotones},
the entanglement $E(\Psi)$ of a pure state $\ket{\Psi}$ can be quantified by the von Neumann
entropy $S(\varrho)$ of the reduced density matrix, \Eq{vNentr}.
Entanglement of formation of a mixed state then follows as the
convex roof~\cite{ben96b}
\be
E(\varrho)=\inf_{\{p_i,\Psi_i\}}\sum_i\ p_i\ E(\Psi_i)\ ,
\hspace{.2cm}\mbox{with}\hspace{.2cm}p_i>0\ ,\hspace{.2cm}
\mbox{s.t.}\hspace{.2cm}
\varrho=\sum_i\ p_i\ \ket{\Psi_i}\bra{\Psi_i}\ .
\label{Eof}
\ee
In arbitrary dimensions the underlying optimisation problem is unsolved -
apart from a few known solutions for particular states~\cite{ter00}.
Only for $2\times 2$ systems an algebraic solution is known for
general states.
For these low dimensional systems, the entanglement of a pure state can be
expressed as a function of its concurrence,
$E(\Psi)={\cal E}(c(\Psi))$, with~\cite{wot98}
\be
{\cal E}(c)=-\sum_{\alpha =-1,1}
\frac12(1+\alpha\sqrt{1-c^2}))\log_2\frac12(1+\alpha\sqrt{1-c^2}))\ .
\ee
The function ${\cal E}(c)$ is monotonically increasing and convex.
Thus, the entanglement of formation can be estimated (with (\ref{roof})) as
\be
E(\varrho)=\inf\ \sum_i\ p_i\ {\cal E}(c_i)\ \ge\
\inf\ {\cal E}\Bigl(\sum_i\ p_i\ c_i\Bigr)=
{\cal E}(c(\varrho))\ .
\ee
Consequently, concurrence provides a lower bound for the entanglement of
formation.
In general, the decomposition that provides the infimum is not unique.
In the present case of $2\times 2$ systems, the manifold on which the
infimum is adopted always contains a set of pure states all of which
have the same concurrence~\cite{wot98}.
For these special decompositions one has
$\sum_ip_i{\cal E}(c_i)={\cal E}\bigl(\sum_ip_ic_i\bigr)$.
Thus, equality holds in the above inequality.
Therefore, in $2\times 2$ systems, one can always express entanglement
of formation in terms of concurrence.
Since an algebraic expression, \Eq{wotinfsol}, for concurrence
is available for arbitrary $2\times 2$ states,
also entanglement of formation can be computed purely algebraically.

\subsection{Higher dimensional systems}
In the following Section we will focus on the concurrence of systems
of arbitrary dimension.
Entanglement of formation is more appealing than concurrence, because
it is believed~\cite{vidal02b,benatti,pomeransky} - though not proven
for general states - to
be additive, whereas it is evident that concurrence is {\em not}
additive.
However, concurrence satisfies several algebraic properties that
provide a basis for good approximations, while it is unknown whether
entanglement of formation can be evaluated as efficiently.
Thus, whereas entanglement of formation may be more appealing from a
rather fundamental point of view, concurrence is more appealing for
pragmatic reasons - it allows for the efficient description of states
even in high dimensional systems, as we will see in the following.

The quantitative estimation of the concurrence of a mixed state is
often achieved by numerical means~\cite{karol99,aud01,tsw03} which
essentially solve a high dimensional optimisation problem when
searching for the minimum that defines the convex roof, \Eq{roof}.
However, such an approach can only provide an upper bound of
concurrence rather than its actual value, since a numerical
optimisation procedure can never guarantee convergence into the global
rather than into a local minimum.
Hence, besides the efficient numerical implementation of
optimisation procedures, it is most desirable to derive a lower bound of
concurrence that can be possibly evaluated in a purely algebraic way, since
the numerical effort increases rapidly with the dimension of the
underlying Hilbert space.
In the present Section, we will derive an approach to characterise
concurrence of general mixed states in arbitrary, finite dimensions. In
particular, we will
\begin{itemize}
\item[-]not only provide a framework for an efficient numerical implementation
  to compute an upper bound of concurrence in Section
 ~\ref{secgrad}, but also
\item[-] propose a generalisation of concurrence
  for multipartite systems in Section~\ref{multi-partite}.
\item[-]Moreover, we will formulate lower bounds of concurrence in
  Section~\ref{seclowb}, some of which can be computed purely algebraically.
\item[-]Finally, in Section~\ref{quasi-pure} we will derive an
  approximation of concurrence that is valid for most states describing current
  experiments, and which can also be evaluated purely algebraically.
\end{itemize}
Note that, when describing higher dimensional systems, our choice of
concurrence as a single scalar quantity can never explore all
correlations inscribed in an arbitrary given state (see also the
discussion at the end of Section~\ref{purbip}). Therefore, the
definition we shall elaborate here will be constructed such as 
to allow to target at different, specific types of correlations, in a
possibly multipartite, higher dimensional quantum system. 

Both, upper and lower bounds of these concurrences, 
will allow to confine their actual values
to finite intervals, providing reliable information
about arbitrary states. Specializing our approach, in
Section~\ref{quasi-pure}, to typical 
experimental requirements, will relax the demand for a completely
general treatment, and will be rewarded by a dramatic speed-up of actual
numerical evaluations, through
a very efficient and easily implemented estimate of concurrence.

Since we have neither an a priori estimate of the tightness of our bounds, nor one for the range of validity of our approximation,
we will compare our estimations from below with the corresponding upper
bound in Section~\ref{seccompare}.
For this purpose, we will use random states, as well as states
under scrutiny in real experiments~\cite{wine00}.

To start with, we have to realize that 
the definition of concurrence given in Section~\ref{secwotconc} only
applies to two-level systems.
Since Bell states used in the original definition in \Eq{wotaltcon},
and the spin flip operation used in \Eq{wotcon}
do not have unique generalisations to higher dimensions,
there is no straightforward generalisation of concurrence
to higher dimensions.
So far, two {\em inequivalent} generalisations for systems comprising
more than two levels have been formulated,
which both coincide with the original one,
if restricted to two-level systems.

\subsubsection{$\Theta$-concurrence}
\label{theta-concurrence}

One possible generalisation is $\Theta${\it -concurrence}~\cite{uhl00}.
The complex conjugation that appears in the {\em bra} in \Eq{wotcon}
can be perceived as an
anti-linear operation.
$\Theta$-concurrence is based on anti-linear operators,
where anti-linearity of an operator $\Theta$ is defined by the property
\be
\Theta\bigl(
\alpha_1\ket{\Psi_1}+\alpha_2\ket{\Psi_2}\bigr)=
\alpha_1^\ast\Theta\ket{\Psi_1}+\alpha_2^\ast\Theta\ket{\Psi_2}\ .
\ee
An anti-linear operator $\Theta$ that is unitary
$\Theta^{-1}=\Theta^\dagger$ and an involution $\Theta=\Theta^{-1}$
is called a {\em conjugation}.
In terms of such a conjugation $\Theta$, one can define
$\Theta$-concurrence $c_\Theta(\psi)$ of a pure (not necessarily normalised)
state $\ket{\psi}$ as
\be
c_\Theta(\psi)=\bigl|\matel{\psi}{\Theta}{\psi}\bigr|\ .
\ee
Of course, $c_\Theta(\psi)$ does not only depend on $\ket{\psi}$, but
also on the choice of $\Theta$.
Thus $\Theta$-concurrence is not a single, uniquely defined quantity,
but rather a family of quantities, depending on the choice of $\Theta$.
In systems larger than two-level systems, no
conjugation $\Theta$ is known such that $c_\Theta(\psi)$ vanishes for all
separable states and is strictly larger than zero for all entangled
states.
However, in the case of two-level systems there is one.
For $\Theta=\sigma_y\otimes\sigma_y\ C_\ast$~\cite{uhl00,WOT01},
with the second Pauli matrix $\sigma_y$ and $C_\ast$ the complex
conjugation in the standard basis (defined after~\Eq{wotcon}),
$c_\Theta(\psi)$ is non-vanishing
if and only if $\ket{\psi}$ is entangled.
For this special choice, $\Theta$-concurrence coincides with regular
concurrence as defined in \Eq{wotcon}.

$\Theta$-concurrence can easily be extended to mixed states
using the concept of convex roofs.
Given a complex symmetric matrix $\tau_\Theta$ with elements
\be
[\tau_\Theta]_{jk}=\matel{\psi_j}{\Theta}{\psi_k}\ ,
\ee
one easily finds for the $\Theta$-concurrence
$c_\Theta(\varrho)=\inf \sum_i p_i c_\Theta(\psi_i)$
of a mixed state $\varrho$:
\be
c_\Theta(\varrho)=
\inf_V\sum_i
\Bigl| \left[V\tau_\Theta V^T\right]_{ii}\Bigr|\ .
\ee
As discussed in Section~\ref{secwotconc}, Eq.~(\ref{wotinfsol}), the infimum
can be expressed as
\be
c_\Theta(\varrho)=\max\left\{\sv_1^\Theta-\sum_{i>1}\sv_i^\Theta,\ 0\right\}\ ,
\ee
with the singular values $\sv_i^\Theta$ of $\tau_\Theta$ in
decreasing order.
Thus, $\Theta$-concurrence can be easily evaluated for arbitrary mixed
states.
However, since, apart from $\Theta=\sigma_y\otimes\sigma_y\ C_\ast$ in
$2\times 2$ systems, no conjugation is known that is positive exactly
for entangled states, it has the disadvantageous property that $c_\Theta$ vanishes for some entangled states.

\subsubsection{I-concurrence}

Alternatively, $I${\it -concurrence}~\cite{run01} is defined in terms
of operators $\inv_1$ and $\inv_2$ acting on ${\cal B}({\cal H}_1$) and
${\cal B}({\cal H}_2)$ as
\be
c_I(\Psi)=\sqrt{\bigl.
\bra{\Psi}(\inv_1\otimes\inv_2\ket{\Psi}\bra{\Psi})\ket{\Psi}
\bigr.}\ .
\label{Iconcurrence}
\ee
The operators $\inv_i$ are required to satisfy the following
properties~\cite{run01}
\begin{itemize}
\item[a)]$\inv_i H=(\inv_i H)^\dagger$ ($i=1,2$),
  for all hermitian operators $H$,
  which ensures that $I$-concurrence is real.
\item[b)]$[\inv_i,{\cal U}]=0$ ($i=1,2$) for all unitary ${\cal U}$,
  which ensures that $I$-concurrence is invariant under local unitary
  transformations.
\item[c)]
  $\bra{\Psi}(\inv_1\otimes\inv_2\ket{\Psi}\bra{\Psi})\ket{\Psi}\ge 0$,
  for all states $\ket{\Psi}$,
  where  equality holds if and only if $\ket{\Psi}$ is separable\ .
\end{itemize}
Up to scaling, there is a unique operator satisfying these
requirements~\cite{run01}, namely
\be
\inv_i(\ket{\varphi_i}\bra{\varphi_i})=
{\mathbbm 1}_{{\cal H}_i}-\ket{\varphi_i}\bra{\varphi_i}\ ,
\ket{\varphi_i}\in{\cal H}_i, \
i=1,2\ ,
\label{orthspace}
\ee
which maps $\ket{\varphi_i}\bra{\varphi_i}$ onto its orthogonal space.
Thus - in contrast to $\Theta$-concurrence - $I$-concurrence is a
quantity that is uniquely defined up to a multiplicative constant.
With \Eq{orthspace}, $I$-concurrence $c_I(\Psi)$ of a pure state
$\ket{\Psi}$ can also be expressed in terms of reduced density
matrices.
With
$\inv_1\otimes\inv_2\prj\Psi=
{\mathbbm 1}-\varrho_1\otimes{\mathbbm 1}_2-
{\mathbbm 1}_1\otimes\varrho_2+\prj\Psi$, one easily obtains
\be
c_I(\varrho)=\sqrt{2-\tr\varrho_1^2-\tr\varrho_2^2}\ .
\label{iconsym}
\ee
As argued right before \Eq{subsys}, the last two terms are equal,
such that there is no need to explicitly distinguish between the two
reduced density matrices.
It therefore became a widespread convention to define concurrence using only one of the
two reduced density matrices
\be
c(\Psi)=\sqrt{2\bigl(1-\tr\varrho_r^2\bigr)}\ ,
\label{iconassym}
\ee
where $\varrho_r$ can be either one.

If we now use the Schmidt form, \Eq{schmidt}, of an arbitrary pure
state $\ket{\Psi}$, its $I$-concurrence reads
\be
c_I(\Psi)=\sqrt{2\sum_{i\neq j}\ \sv_i\sv_j}\ ,
\ee
and it is easily verified that $I$-concurrence coincides with the
original definition given in \Eq{wotcon}, for two-level systems.
Note that $I$-concurrence cannot exceed a given maximum value,
\begin{equation}
c_I(\Psi)\leq\sqrt{2(1-1/\dss)}\ ,
\label{maxconcI}
\end{equation}
with $\dss$ the dimension of the smallest subsystem.

\subsection{Representation in product spaces\label{product-spaces}}
To start with, we would like to represent the definition of
concurrence, \Eq{Iconcurrence}, in a different form.
Whereas \Eq{Iconcurrence} allows for a nice interpretation of the
entanglement of pure states $\ket{\Psi}$ in terms of the degree of
mixing of the reduced density matrices, it is not very suitable for
the evaluation of the convex roof~(\ref{roof}), if we are dealing with
mixed states.
What we are looking for instead, is a {\em linear} operator $a$ such that all
expectation values $\matel{\Psi}{a}{\Psi}$ with respect to separable pure
states vanish,
whereas they are strictly positive for entangled pure states.
However, such an operator does not exist.
In the case of two-level systems, one could circumvent this problem by
using $\bra{\Psi^\ast}$ instead of $\bra{\Psi}$ (see \Eq{wotcon}).
However, this trick does not work properly in higher dimensional
systems, as discussed in \ref{theta-concurrence}.

In order to obtain a quantity that unambiguously detects all entangled
states in arbitrary dimensions, we can follow a different way. We
consider expectation values with respect to {\em two copies} of
the pure state under investigation,
\begin{figure}
\epsfig{file=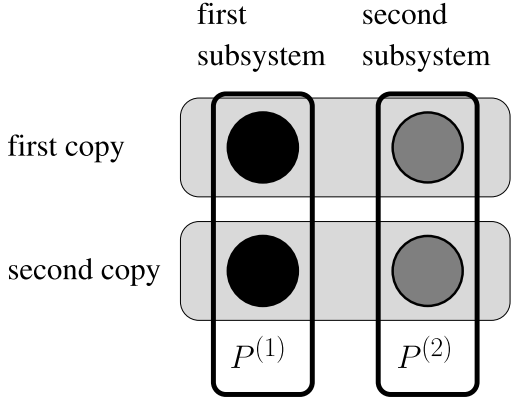,width=.5\textwidth,angle=0}
\caption{An expectation value $\matel{\Psi}{a}{\Psi}$ with respect to a
single copy of a given state $\ket{\Psi}$ can not discriminate
separable from entangled states.
However, this is possible, if expectation values with respect to two
copies are used.
The grey boxes symbolise the two copies of the state $\ket{\Psi}$,
with first subsystem (black) and second subsystem (grey). This scheme
defines the structure of an operator acting on two copies of the
single subsystems, such that the corresponding expectation values
vanish exactly if the state $\ket{\Psi}$ is separable.}
\label{copies-1}
\end{figure}
\be
c(\Psi)=
\sqrt{\bigl.\bra\Psi\otimes\bra\Psi\ A\
\ket\Psi\otimes\ket\Psi\bigr.}\ ,
\label{concurrence-A}
\ee
where $A$ is acting on ${\cal H}\otimes{\cal H}$, {\it i.e} on
${\cal H}_1\otimes{\cal H}_2\otimes{\cal H}_1\otimes{\cal H}_2$.
Of course, one needs to require that \Eq{concurrence-A} vanishes for any
separable state
$\ket{\Psi}=\ket{\varphi}\otimes\ket{\phi}$.
The simplest possibility could be an operator $A$ that decomposes into a part
acting only on the copies of the first subsystem and a second part acting on
the copies of the second subsystem.
Not necessarily the unique, but a good choice for $A$ are projectors $P_-^{(k)}$
$(k=1,2)$ onto antisymmetric subspaces ${\cal H}_-^k$ of the space
${\cal H}_k\otimes{\cal H}_k$.
They contain all states that acquire a phase
shift of $\pi$ under the exchange of the two copies of $\cal H_k$.
Thus, any state $\ket{\psi_-}\in{\cal H}_-^k$ can be expressed as
$\ket{\psi_-}=\sum_{ij}\psi_{ij}(\ket{ij}-\ket{ji})$,
where the states $\{\ket{i}\}$ and $\{\ket{j}\}$ form an arbitrary basis of $\cal H_k$.

Indeed, since the two-fold copy $\ket\varphi\otimes\ket\varphi$ of a state is
a symmetric object - it remains invariant under an exchange of the copies -
the expectation value
$\bra\varphi\otimes\bra\varphi P_-^{(1)} \ket\varphi\otimes\ket\varphi$
vanishes for any state $\ket\varphi\in{\cal H}_1$, and the same holds true for
the analogous expression for ${\cal H}_2$.

Now, one can define $A:=4P_a$ in \Eq{concurrence-A},
with the projector $P_a$ onto the space
spanned by the states in
${\cal H}_1\otimes{\cal H}_2\otimes{\cal H}_1\otimes{\cal H}_2$
that are antisymmetric both with respect to an exchange of the two copies of
${\cal H}_1$, and with respect to an exchange of the two copies of
${\cal H}_2$.
The concurrence in~\Eq{concurrence-A} then necessarily vanishes for
separable states.
The prefactor $4$ is just a normalisation chosen such that the
concurrence ranges from $0$ to $1$ for two-level systems.
With this normalisation, the present definition is indeed equivalent
to~\Eq{iconassym}.

One may find it easier to interprete $A$ in terms of the two copies of
the single subspaces ${\cal H}_1$ and ${\cal H}_2$.
For this purpose, let's identify
${\cal H}_1\otimes{\cal H}_2\otimes{\cal H}_1\otimes{\cal H}_2$ and
${\cal H}_1\otimes{\cal H}_1\otimes{\cal H}_2\otimes{\cal H}_2$.
Then - as illustrated in Fig.~\ref{copies-1} - $A$ is just the tensor product of the two projectors onto the two involved anti-symmetric
subspaces:
\begin{equation}
A=4P_-^{(1)}\otimes P_-^{(2)}\ .
\label{Abip}
\end{equation}
If a state $\ket{\Psi}$ is separable, the expectation value
$\bra\Psi\otimes\bra\Psi A\ket\Psi\otimes\ket\Psi$
factorises into the product of the analogous expressions corresponding
to the single subsystems
\be
\bra\Psi\otimes\bra\Psi A\ket\Psi\otimes\ket\Psi=4
\bra\varphi\otimes\bra\varphi P_-^{(1)}\ket\varphi\otimes\ket\varphi
\bra\phi\otimes\bra\phi P_-^{(2)}\ket\phi\otimes\ket\phi\ ,
\ee
and vanishes because both, $\ket\varphi\otimes\ket\varphi$ and
$\ket\phi\otimes\ket\phi$, are symmetric.
However, for an entangled state
$\ket{\Psi}=\sum_i\sqrt{\lambda_i}\ket{\sch_i}_1\otimes\ket{\sch_i}_2$
- for convenience represented in its Schmidt decomposition, \Eq{schmidt} -
the two-fold copy $\ket\Psi\otimes\ket\Psi$ is not symmetric under exchange of the copies of any of the subsystems.
Thus, $\ket\Psi\otimes\ket\Psi$ necessarily has some anti-symmetric
part, and the expectation value of $A$ with respect to this two-fold
copy of $\ket\Psi$ is strictly positive.

Now, with this definition of $A$, one can also rephrase the
concurrence of a mixed state $\varrho$ in terms of subnormalised
states, Eq.~(\ref{subnormalised}), as
\be
c(\varrho)=\inf_{\{\ket{\psi_i}\}}
\sum_i\sqrt{
\bra{\psi_i}\otimes\bra{\psi_i}\ A\
\ket{\psi_i}\otimes\ket{\psi_i}}\ .
\ee
If one makes use of the prescription (\ref{ensemble}) to characterise
all decompositions of $\varrho$ into pure states,
it reveals useful to define a
tensor $\cal A$ that contains the elements of $A$ evaluated with the
subnormalised states $\{\ket{\phi_j}\}$
\be
{\cal A}_{jk}^{lm}=
\bra{\phi_l}\otimes\bra{\phi_m}\ A\
\ket{\phi_j}\otimes\ket{\phi_k}\ .
\label{elements-A}
\ee
One can now easily rewrite $c(\varrho)$ in the closed expression
\be
c(\varrho)=\inf_V\sum_i\sqrt{
\bigl[V\otimes V\ {\cal A}\ V^\dagger\otimes V^\dagger\bigr]_{ii}^{ii}}\ ,
\label{mixed-concurrence}
\ee
where the optimisation is to be performed over all left unitaries $V$.

\subsubsection{Symmetries of $A$\label{symmetries}}

We required $A$ to be anti-symmetric with respect to the exchange of the two
copies of ${\cal H}_1$ as well as of ${\cal H}_2$.
Though, this choice is not unique.
It would have been sufficient to require only one of these symmetries.
Here, we will briefly sketch the consequences of specific choice of $A$.
In Section~\ref{two-level-systems} and later on in Section~\ref{seclowb}, it
will become apparent why it is of importance to require both symmetries.

With our above definition, Eq.~(\ref{Abip}), of $A$,
the elements of $\cal A$
\bqa
{\cal A}_{jk}^{lm}&=&
\tr\bigl[\ket{\phi_j}\ovl{\phi_l}{\phi_k}\bra{\phi_m}\bigr]-
\tr_1\Bigl[
\tr_2\bigl[\ket{\phi_j}\bra{\phi_l}\bigr]
\tr_2\bigl[\ket{\phi_k}\bra{\phi_m}\bigr]\Bigr]-\nonumber\\
&&
\tr_2\Bigl[
\tr_1\bigl[\ket{\phi_j}\bra{\phi_l}\bigr]
\tr_1\bigl[\ket{\phi_k}\bra{\phi_m}\bigr]\Bigr]+
\tr\bigl[\ket{\phi_j}\bra{\phi_l}\bigr]
\tr\bigl[\ket{\phi_k}\bra{\phi_m}\bigr]
\eqa
contain partial traces over both subsystems, in a balanced way,
what would not have been the case for other choices.
{\it E.g.}, for a projector ${\tilde A}$ onto the space spanned by all the states that are
antisymmetric with respect only to the exchange of ${\cal H}_1$,
the analogous expression reads
\be
\tilde {\cal A}_{jk}^{lm}
=2\left(
\tr\bigl[\ket{\phi_j}\ovl{\phi_l}{\phi_k}\bra{\phi_m}\bigr]-
\tr_1\Bigl[
\tr_2\bigl[\ket{\phi_j}\bra{\phi_l}\bigr]
\tr_2\bigl[\ket{\phi_k}\bra{\phi_m}\bigr]\Bigr]\right)\ .
\ee
Whereas the symmetric treatment of both subsystems does not have any
practical consequences, and is of rather aesthetical character,
there is a much more crucial property:
${\cal A}$ is invariant under exchanges of the co- or contra-variant
indices, ${\cal A}_{jk}^{lm}={\cal A}_{kj}^{lm}={\cal A}_{jk}^{ml}$.
This symmetry will turn out to be the crucial ingredient required for
the approximations to be discussed in Section~\ref{seclowb}.

\subsubsection{Two-level systems\label{two-level-systems}}

In the case of two-level systems, there is only one anti-symmetric
state, namely $\ket{01}-\ket{10}$.
Therefore, the projectors onto the
anti-symmetric subspaces ${\cal H}_-^1$ and
${\cal H}_-^2$
 have only one non-vanishing eigenvalue.
This, in turn, also holds true for $A$ which reads
$A=\prj\chi$, with
\be
\ket{\chi}=
\ket{0011}-\ket{0110}-\ket{1001}+\ket{1100}\ .
\ee
Consequently, ${\cal A}$ can be expanded in terms of a single matrix
$\tau$ as
\be
{\cal A}_{jk}^{lm}=\tau_{lm}^\ast\tau_{jk}\ ,
\hspace{.5cm}\mbox{with}\hspace{.5cm}
\tau_{jk}=\trol{\chi}{\phi_j}{\phi_k}\ .
\ee
Due to the symmetry of $A$ discussed above in Section~\ref{symmetries},
$\tau$ is indeed symmetric, {\it i.e.} satisfies a crucial
precondition for the generalisation to mixed states.
Expressed in terms of $\tau$, \Eq{mixed-concurrence} now simplifies to
\be
c(\varrho)=\inf_V\sum_i\sqrt{
\left[V^\ast \tau^\ast V^\dagger\right]_{ii}
\left[V\tau V^T\right]_{ii}}=
\inf_V\sum_i\left|\left[V\tau V^T\right]_{ii}\right|\ .
\label{two-level-concurrence}
\ee
Due to the symmetry of $\tau$, the infimum is exactly given in terms of
the singular values of $\tau$, as discussed in the context of \Eq{wotinf}.

The relation between the present and the original approach to
concurrence~\cite{wot98,wot97} becomes apparent with the observation
that the matrix $\tau$ defined here coincides with that
introduced in Eq.~(\ref{wotinf})
\be
\tau_{jk}=
\trol{\chi}{\phi_j}{\phi_k}=
\matel{\phi_j^\ast}{\sigma_y\otimes\sigma_y}{\phi_k}\ ,
\ee
in the $2\times 2$ case. Thus, the original approach to concurrence is naturally embedded in the present,
more general framework.

\subsubsection{Higher dimensional systems}
\label{sec-higher-dim-sys}

The formalism in terms of projectors onto
anti-symmetric subspaces can also be used to formulate a
generalisation of concurrence to systems of arbitrary dimensions.
For an $n_1\times n_2$-dimensional system, $A$ has
$m=n_1(n_1-1)n_2(n_2-1)/4$ non-vanishing eigenvalues - $n_1(n_1-1)/2$
corresponding to the antisymmetric subspace of
${\cal H}_1\otimes{\cal H}_1$, and analogously for ${\cal H}_2\otimes{\cal H}_2$.
Thus, $A$ cannot be expressed with the help of a single eigenvector
anymore, but with a finite sum
\begin{equation}
A=\sum_{\alpha=1}^m\prj{\chi_\alpha}\ ,
\end{equation}
which contains $n_1(n_1-1)n_2(n_2-1)/4$ non-vanishing terms.
Due to the symmetries
${\cal A}_{jk}^{lm}={\cal A}_{kj}^{lm}={\cal A}_{jk}^{ml}$
which we already observed in Section~\ref{symmetries}, all $T^\alpha$
with the elements
\be
T_{jk}^{\alpha}=\trol{\chi_{\alpha}}{\phi_j}{\phi_k}\ ,
\alpha=1,\hdots,m=n_1(n_1-1)n_2(n_2-1)/4\ ,
\label{definition-T}
\ee
are symmetric.
Furthermore, the eigenvectors of $A$ still carry an
undetermined phase factor $e^{i\varphi_\alpha}$.
Whereas these free phases usually do not matter, they provide an additional
freedom which we shall exploit in Section~\ref{seclowb}.
Therefore, we explicitly account for the free phases $\varphi_\alpha$,
and \Eq{mixed-concurrence} consequently reads
\be
c(\varrho)=\inf_V\sum_i\sqrt{
\sum_\alpha \left|\left[VT^\alpha e^{i\varphi_\alpha} V^T\right]_{ii}\right|^2}\ .
\label{concurrence-T}
\ee
There are two crucial differences as compared to
\Eq{two-level-concurrence} that hamper finding the exact infimum:
First, the square and the square root in Eq.~(\ref{concurrence-T})
lead to non-linear expressions in the $T^\alpha$,
and, second, the already mentioned fact that the several distinct
symmetric matrices $T^\alpha$ {\em cannot}, in general, be diagonalised simultaneously.

One of the earliest generalisations of concurrence to higher
dimensional systems that does not lead to the non-linear behaviour, as it
appears in \Eq{concurrence-T}, is the {\em concurrence vector}.
Although the original definition~\cite{aud01} is slightly different,
we will describe it here in terms of the projectors of \Eq{Abip}, in order
to highlight the similarities with our own approach.
Each $\ket{\chi_{\alpha}}$ inherits a negative pre-factor under the exchange of
the two copies of either ${\cal H}_1$ or ${\cal H}_2$.
The product $\ket{\Psi_s}\otimes\ket{\Psi_s}$ of an arbitrary
{\em separable} state is invariant under such transformations.
Since the overlap
\be
{\cal C}_{\alpha}(\Psi)=\trol{\chi_{\alpha}}{\Psi}{\Psi}
\ee
is invariant under the exchange of the two  copies, it necessarily
needs to vanish for any $\ket{\chi_{\alpha}}$ and any separable state
  $\ket{\Psi}=\ket{\Psi_s}\otimes\ket{\Psi_s}$.
In contrast, whenever ${\cal C}_{\alpha}(\Psi)$ is positive, the state $\ket{\Psi}$
necessarily needs to be entangled.
The inverse implication is a bit more involved -
$\ket{\Psi}$ is separable, if ${\cal C}_{\alpha}(\Psi)=0$, for
$\alpha=1,\hdots,n_1(n_1-1)n_2(n_2-1)/4$.

The generalisation of the above for mixed states is now straightforward and
analogous to the case of two-level systems discussed
in Section~\ref{two-level-systems}.
A given state $\varrho$ is separable if and only if there is a
left-unitary transformation $V$ such that all elements
\be
{\cal C}_{\alpha}=\sum_j\left|\left[VT^{\alpha}V^T\right]_{jj}\right|\ ,
\ee
of the {\em concurrence vector} ${\cal C}$ vanish. However,
this -- necessary and sufficient -- separability criterion is, in
general, difficult to evaluate -- since, again, in general the
matrices $T^{\alpha}$ cannot
be diagonalised simultaneously.
Nevertheless, it establishes the basis for some operational, though only necessary
separability criteria.
It implies that a given state is entangled if the singular values
$\sv_j^{({\alpha})}$ of one matrix $T^{\alpha}$ satisfy
$\sv_1^{({\alpha})}-\sum_j\sv_j^{({\alpha})}>0$.
Another -- in general stronger -- criterion is obtained with the help
of linear combinations $\sum_{\alpha}z_{\alpha}T^{\alpha}$ of all matrices $T^{\alpha}$, with
complex pre-factors $z_{\alpha}$.
For a suitably chosen set $\{z_{\alpha}\}$, the expression $\sv_1-\sum_j\sv_j$
can be significantly larger than the corresponding expression
for a single matrix $T^{\alpha}$ \cite{aud01}.

Concurrence vector is not a measure and as such does not provide an
adequate tool for our aims, {\it i.e.} the quantitative
characterisation of temporally evolving entanglement.
Though, in section~\ref{seclowb} we will show that - despite its
non-linearity - \Eq{concurrence-T} can be used to derive some means
for such a quantification.

\subsubsection{Gradient\label{secgrad}}

Before we focus on this, however, let's discuss how to assess
concurrence numerically.
As mentioned in Section~\ref{mixedstates}, the cardinality of
the ensemble that realises the infimum in \Eq{roof} can exceed the
rank $r(\varrho)$, though is bounded by $r^2(\varrho)$~\cite{uhl98}.
This is the cause of the appearance of a rectangular, left-unitary matrix $V$
instead of a quadratic, unitary matrix $U$ in \Eq{mixed-concurrence}.
In the present section, however, matrices of the latter type are more
convenient.
Therefore, we will fix the cardinality of the considered ensembles.
If it turns out that the assumed cardinality is not large enough,
one can always increase it by adding some null-vectors to the ensemble.

According to \Eq{mixed-concurrence},
the concurrence of a mixed state $\varrho$ is given by
\be
c(\varrho)=\inf_{U}{\cal C}(U)\ ,
\hspace{.5cm}\mbox{with}\hspace{.5cm}
{\cal C}(U)=\sum_i
\sqrt{\left[U\otimes U {\cal A}\
U^\dagger\otimes U^\dagger\right]_{ii}^{ii}}\ .
\label{upbound}
\ee
If one considers an infinitesimal transformation
$dU={\mathbbm 1}+i\epsilon K$
, with $\epsilon$ infinitesimally small, and
uses the symmetry of ${\cal A}$ with respect to an exchange of co- and
contravariant indices, this can be written as
\be
{\cal C}(dU)\simeq\sum_i
\sqrt{A_{ii}^{ii}}+\frac{i\epsilon}{\sqrt{A_{ii}^{ii}}}\left[
K\otimes{\mathbbm 1}\ A\ -
A\ K\otimes{\mathbbm 1}
\right]_{ii}^{ii}\ ,
\label{inftrans}
\ee
where the expansion
$\sqrt{a+\varepsilon}\simeq\sqrt{a}+\varepsilon/(2\sqrt{a})$ of the square
root function was employed.
This allows to rephrase ${\cal C}(dU)$ as
\be
{\cal C}(dU)=\sum_i\sqrt{A_{ii}^{ii}}+
\varepsilon\ \sum_{ij}K_{ij}G_{ji}\ ,
\hspace{.5cm}\mbox{with}\hspace{.5cm}
G_{ji}=i\left(
\frac{A_{ji}^{ii}}{\sqrt{\big.A_{ii}^{ii}}}-
\frac{A_{jj}^{ij}}{\sqrt{A_{jj}^{jj}}}\right)\ .
\label{gradient-A}
\ee
Since the square root is not analytic in
the origin,
one has to take care that $A_{ii}^{ii}$ be non-vanishing for all $i$.
This is the case if and only if there are only non-separable pure
states $\ket{\Psi_i}$ in the decomposition of $\varrho$,
since, by \Eq{concurrence-A} and \Eq{elements-A}, the elements
$A_{ii}^{ii}$ are the squares of the concurrences of the states $\ket{\Psi_i}$.
Though, even if the denominators in \Eq{gradient-A} vanish, the
numerators behave accordingly~\cite{floPhD},
such that \Eq{gradient-A} is indeed always well defined.

The hermitian matrix $G$ in \Eq{gradient-A} can be considered as a gradient,
since the increment of ${\cal C}$ reads
\be
{\cal C}(dU)-{\cal C}({\mathbbm 1})=\varepsilon\ \tr\left[KG\right]\ ,
\ee
what is just a Hilbert-Schmidt scalar product~\cite{horn85}.
Accordingly, the direction of steepest descent of $\cal C$ is given by
$-G$.
Minima of ${\cal C}$ can therefore be found by
repeated application of the transformation $dU=\exp(-i\epsilon G)$.
However, also more refined methods can be used, such as the
{\it conjugate-gradient minimisation}~\cite{aud01,matcomp}.
In this method, the iteration is not performed exactly along the
current gradient $G_n$, but it takes into account corrections
that ensure that the previous iteration along the gradient
$G_n$ is not reversed.
More explicitely the iteration is performed along
\bqa
\tilde G_n=G_i-\gamma G_{n-1}\ ,
&\mbox{with}&
\gamma=\frac
{\tr\left[\left(G_n-\tau G_{n-1}\right)G_n\right]}
{\tr\left[G_{n-1}^2\right]}\ ,\nonumber\\
&\mbox{and}&
\tau G_{n-1}=
e^{i\frac\varepsilon 2\tilde G_{n-1}}G_{n-1}
e^{-i\frac\varepsilon 2\tilde G_{n-1}}\ .
\eqa
However, there is in general no a priori information available on whether the
solution reached by that procedure is a local or a global minimum.
While one may start the iteration with different initial conditions
parametrised by $U$, in order to get a better intuition on the effective
``landscape'' defining the optimisation problem at hand, this uncertainty
persists, and the more so the higher the dimension of the parameter space over
which the optimisation is carried out.

\subsubsection{Multi-partite systems\label{multi-partite}}

Since lately several experimental groups
\cite{wein00,hae,wine00,panPRL01,eiblPRL04,roos,rauschSCI00,zhaoNAT04}
systematically investigate quantum correlations in
multipartite systems, {\it i.e.}, systems with more than two
subsystems,
a quantitative description of multipartite entanglement is highly
desirable.
Multipartite systems keep room for distinct classes of quantum states:
Even for a pure state $\ket{\Psi}$ of a tri-partite two-level system
more than a single scalar quantity is required for complete characterisation
of all inscribed quantum correlations.

In bipartite systems, any state can be prepared using LOCC only,
starting with a distinguished, maximally entangled state (see Section
\ref{purbip}).
This is no longer true in multipartite systems -
where inequivalent kinds of multipartite entanglement exist~\cite{duer00}.
Consider for example a {\it Greenberger-Horne-Zeilinger state} (GHZ-state)~\cite{GHZ}
\be
\ket{\Psi_{\rm{GHZ}}}=\frac{1}{\sqrt 2}\bigl(\ket{000}+\ket{111}\bigr)\ ,
\ee
and a W-state~\cite{duer00}
\be
\ket{\Psi_{\rm{W}}}=\frac{1}{\sqrt 3}
\bigl(\ket{001}+\ket{010}+\ket{100}\bigr)\ .
\ee
Both states contain fundamentally different correlations,
such that none of the two can be created from the other one by LOCC
alone~\cite{duer00}.
Thus, one cannot expect that a single scalar quantity can completely describe
$\nss$-particle correlations (with $p>2$).
\begin{figure}
\epsfig{file=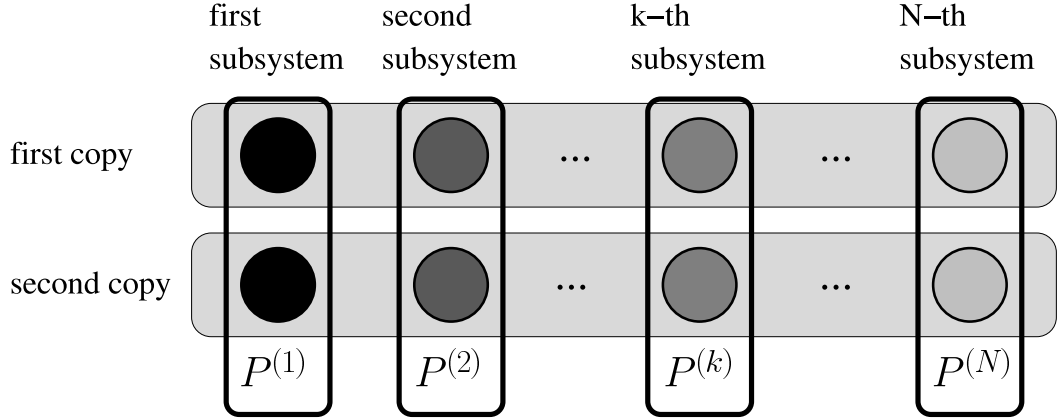,width=1.0\textwidth,angle=0}
\caption{
The concurrence of bipartite states is defined in terms of two
operators, each of which is associated with two copies of one
subsystems, as displayed in Fig.~\ref{copies-1}.
This concept can straightforwardly be generalised to multipartite
systems.
For each two copies of one subsystems, there is one operator $P^{(k)}$.}
\label{copies-2}
\end{figure}

Hence, how can we generalise the concept of concurrence to
multipartite systems?
The definition of concurrence in \Eq{iconassym} is not very
suggestive for generalisations:
While the degree of mixing of the reduced density matrix has an unambiguous
interpretation for bipartite systems, its meaning is unclear in the multipartite case.
Though, \Eq{concurrence-A} has a rather obvious multipartite formulation, what implies a
generalisation of concurrence for multipartite systems that
can describe at least some of the correlations we are seeking for.
A generalisation of concurrence~\cite{wot00} for
tri-partite two-level systems is already available -- it characterises all tri-partite
correlations of pure states. In the following, we present an even more general
framework applicable to
systems with
an arbitrary number of subsystems, for pure {\em and} mixed states.

Similarly to the case of bipartite concurrence, also its multipartite
generalisations can be
defined in terms of an operator $A$ acting on the tensor
product of ${\cal H}$ with itself.
The only difference being that ${\cal H}$ is the tensor product of more
than two factor spaces ${\cal H}_i$.
Any tensor product of projectors onto symmetric
and anti-symmetric sub-spaces is positive semi-definite,
and invariant under local unitary transformations, see Fig.~\ref{copies-2}.
However, not all such operators finally lead to tensors ${\cal A}$ with the
desired invariance under exchange of co- or contravariant indices.
This symmetry is only valid for products of an even number of projectors
$P_-^{(i)}$ onto anti-symmetric subspaces, possibly multiplied with some
projectors $P_+^{(i)}$ onto symmetric subspaces.
On the other hand, all expectation values of the type appearing on the right
hand side of Eq.~(\ref{concurrence-A}), with $A$ a
product of an odd number of projectors onto
anti-symmetric subspaces, vanish anyway, for arbitrary states.

Thus, we define $N$-partite concurrence as in \Eq{concurrence-A},
{\it i.e.}
\be
c(\Psi)=
\sqrt{\bigl.\bra\Psi\otimes\bra\Psi\ A\
\ket\Psi\otimes\ket\Psi\bigr.}\ ,
\ee
in terms of a  sum of
direct products of projectors onto symmetric and anti-symmetric
subspaces
\be
A=\sum_{{\cal V}{\{s_i=\pm\}}\atop{\prod_{i=1}^Ns_i=+}} p_{\{s_i\}}
\bigotimes_{j=1}^N P_{s_j}^{(j)}\ ,\
p_{\{s_i\}}\ge 0\ .
\label{defA}
\ee
Here, ${\cal V}{\{s_i=\pm\}}$ represents all possible variations of an
$N$-string of the symbols $+$ and $-$,
and the summation is restricted to contributions with an even number of
projectors onto anti-symmetric subspaces.

Note that, because of the continuous parametrisation in terms of the
$p_{\{s_i\}}$, Eq.~(\ref{defA}) actually defines a continuous family of
multipartite concurrences. While the intuitive interpretation of the
concurrence defined by an arbitrary choice of the $p_{\{s_i\}}$ so far remains
an open problem, there are some specific choices of the $p_{\{s_i\}}$ that have immediate
applications in the characterisation of the multipartite entangled states dealt
with in Section \ref{secdyn}. Let's take all prefactors $p_{\{s_i\}}$ in Eq. (\ref{defA}) equal,
with the only exception of setting $p_{+\hdots +}=0$.
As in Eq.~(\ref{Abip}), there is some freedom in the normalisation.
Once again we set $p_{\{s_i\}}=4$ - just to be consistent with Eq.~(\ref{Abip}).
The concurrence $c_N$, defined with these prefactors for systems with
an arbitrary number $N$ of subsystems, can be
expressed in terms of the reduced density matrices $\varrho_i$
~\cite{floandre} as
\be
c_N(\Psi)=2^{1-\frac{N}{2}}
\sqrt{(2^N-2)\ovl{\Psi}{\Psi}^2- \sum_i\tr\varrho_i^2}\ .
\label{cN}
\ee
The index $i$ runs over all $(2^N-2)$ nontrivial subsets of
an $N$-particle system.
Obviously, $c_N$ vanishes only for completely separable $N$-particle states,
since all reduced density matrices are simultaneously pure only for these states. What is
less obvious but at least equally important is that $c_N$
allows to compare the entanglement
of $N$- and $(N-1)$-partite states:
for any state $\ket{\Psi}$ that factorises into a product of a
one-particle state and a $(N-1)$-partite remainder $\ket{\varPhi}$,
$c_N(\Psi)$ reduces to $c_{N\mbox{-}1}(\varPhi)$. Thus, $c_N$ is perfectly
suited to investigate the scaling properties of multipartite entanglement with
the system size $N$, what we shall explore in Section \ref{dynmultpart}.

Further entanglement-properties can be addressed by other choices of
the $p_{\{s_i\}}$ in Eq.~(\ref{defA}): for example, bi-separability of
tri-partite states can be characterised by $c_3^{(23)}$, with
$A_3^{(23)}=4P_+^{(1)}\otimes P_-^{(2)}\otimes P_-^{(3)}$, and, analogously, by $c_3^{(13)}$, with
${A}_3^{(13)}=4P_-^{(1)}\otimes P_+^{(2)}\otimes P_-^{(3)}$,
or by $c_3^{(12)}$, with
${A}_3^{(12)}=4P_-^{(1)}\otimes P_-^{(2)}\otimes P_+^{(3)}$.
The subscript represents the number of subsystems as in Eq.~(\ref{cN})
while the superscript stands for the subsets in which quantum
correlations are measured\footnote{Since in Eq.~(\ref{cN}) we take
  into account all possible subsets, or partitions, we dropped there the use of the superscript.}.
Consider $\ket{\Psi}=\ket{\varPhi_{12}}\otimes\ket{\varphi_3}$ as an exemplary
state.
The choice $c_3^{(12)}$ quantifies the bipartite concurrence of the
entangled part $\ket{\varPhi_{12}}$ of $\ket{\Psi}$, {\it i.e.},
$c_3^{(12)}(\Psi)=c(\varPhi_{12})$,
whereas $c^{(23)}_3$ and $c^{(13)}_3$ vanish
identically for this bi-separable state.
This selectivity with respect to correlations between specific
subgroups of subsystems also hold for bi-separable mixed states - guaranteed by
the construction (\ref{roof}) as convex roof.

Another interesting quantity emerges for four-partite systems:
Since the number of subsystems is even,
there is a term
${A}_4^{(1234)}=4P_-^{(1)}\otimes P_-^{(2)}\otimes P_-^{(3)}\otimes P_-^{(4)}$
that does not contain any factor $P_+^{(i)}$.
The corresponding concurrence vanishes for all states that do not
contain proper four-partite correlations,
{\it i.e.}, bi-separable and tri-separable states.
In particular, for a GHZ state, $c_4^{(1234)}(\Psi_{\rm
  GHZ})=2\sqrt{\sum_{i\neq j}\lambda_i\lambda_j}$
while $c_4^{(1234)}$ vanishes for $W$-states --
similarly like for tri-partite systems where $W$-states bear only bipartite correlations
.
In the specific case of two-level systems, this particular choice of $c_4^{(1234)}$ measures the potential of a given state for
multi-particle teleportation~\cite{rigolin}.

\section{Lower bounds \label{seclowb}}
In the case of bipartite two-level systems, we were able to evaluate the
concurrence of arbitrary mixed states $\varrho$ exactly.
This was possible because ${\cal A}$ (see \Eq{elements-A}) was of rank one.
For higher dimensional systems ${\cal A}$ typically is of higher rank,
such that \Eq{concurrence-T} exhibits two complications with respect to
\Eq{two-level-concurrence}:
additional non-linearities, and 
different matrices $T^\alpha$ which, in general, {\em cannot} be diagonalised
simultaneously. 
These two properties have so far prevented the derivation of an
explicit solution of the optimisation problem formulated in
\Eq{concurrence-T}.
However, as we will show now, concurrence can be bounded from below, by some 
suitable approximations~\cite{flo03,floqp}.

First, the Cauchy-Schwarz inequality~\cite{inequal}
\be
\Bigl(\sum_\alpha x_\alpha^2\Bigr)^\frac12
\Bigl(\sum_\alpha y_\alpha^2\Bigr)^\frac12\ge
\sum_\alpha x_\alpha y_\alpha
\label{CS}
\ee
allows to linearise \Eq{concurrence-T}. With
$x_\alpha:=\left|\bigl[VT^\alpha e^{i\varphi_\alpha}V^T\bigr]_{ii}\right|$,
we conclude that
\be
c(\varrho)\ge\inf_V\sum_{i=1}^N\sum_{\alpha=1}^my_\alpha
\left|\Bigl[VT^\alpha e^{i\varphi_\alpha}V^T\Bigr]_{ii}\right| ,
\hspace{.5cm}\mbox{with}\hspace{.5cm}
\sum_\alpha y_\alpha^2=1\ ,
\label{nachCS}
\ee
where we introduced some auxiliary real parameters $y_\alpha$.
However, the different 
matrices $T^\alpha$
still cause trouble finding the desired infimum. Here,
the triangle inequality
\be
\sum_\alpha\bigl|z_\alpha\bigl|\ge\bigl|\sum_\alpha z_\alpha\bigl|\ ,
\label{inequ2}
\ee
valid for arbitrary complex numbers $z_\alpha$, allows to circumvent 
this problem:
For $z_\alpha:=y_\alpha
\bigl[VT^\alpha e^{i\varphi_\alpha}V^T\bigr]_{ii}$, and
$y_\alpha\ge 0$,
one obtains
\be
c(\varrho)\ge\inf_V\sum_{i=1}^N
\left|\Bigl[
V\Bigl(\sum_{\alpha=1}^m y_\alpha T^\alpha e^{i\varphi_\alpha}\Bigr)V^T
\Bigr]_{ii}\right|\ ,
\label{lowb}
\ee
an expression for which the infimum is given analytically by \Eq{infimum}. 
Our final expression for a 
lower bound of the concurrence therefore 
reads 
\be
c(\varrho)\ge\max\left\{\sv_1-\sum_{i>1}\sv_i,\ 0\right\}\ ,
\label{bound}
\ee
with the singular values $\sv_j$ of
\begin{equation}
{\cal T}=\sum_\alpha Z_\alpha T^\alpha,\
Z_\alpha=y_\alpha e^{i\varphi_\alpha},\ {\rm and}\
{\sum_\alpha|Z_\alpha|^2=1}\ .
\label{freeparbound}
\end{equation}

The bound in \Eq{bound} still depends on the choice of the $Z_\alpha$,
what allows to tighten the estimate.
Thus, one is left with an optimisation problem on an
$2m$-dimensional parameter space~\cite{flo03},
where $m$ is the number of matrices $T^\alpha$ in Eq.~(\ref{definition-T}).
Note that the constraint $\sum_\alpha|Z_\alpha|^2=1$ is by far simpler to
implement than 
left unitarity, 
\Eq{ensemble},
since it is easily
parametrised.
Moreover, the dimension $m$ of optimisation space is significantly
reduced
as compared to the dimension $n_1^3 n_2^3$ of the original
optimisation problem defined by Eqs.~(\ref{roof}) and 
(\ref{ensemble}).

\subsection{Purely algebraic bounds\label{algebraic-bound}}
The lower bound of \Eq{lowb} 
was obtained by the replacement of several
matrices $T^\alpha$ by a single suitably chosen
${\cal T}=\sum_{\alpha} Z_{\alpha} T^{\alpha} $.
So far, there is no clear prescription of how to chose the expansion
coefficients $Z_\alpha$,
which is partially due to the fact that the $T^\alpha$ are determined
only up to degeneracy.
They are constructed with the help of the eigenvectors
$\ket{\chi_\alpha}$ of the projector $A$, what specifies an
eigenspace,
but does not distinguish states within these subspaces.

One way 
to get rid of this 
ambiguity is to diagonalise
${\cal A}$ instead of $A$.
For a typical state $\varrho$, ${\cal A}$
will 
have no degenerate eigenvalues, and thus
has uniquely defined eigenvectors $T^\alpha$.
One can then provide $m$ different lower bounds of concurrence
  directly calculating the singular values $\sv_j$ in
  Eq.~(\ref{bound}) for
\be
{\cal T}=T^\alpha,\ \alpha=1, \hdots m.
\label{T-pab}
\ee
In Section~\ref{seccompare} we will see that one of these bounds alone may already yield a satisfactory approximation to the optimised lower bound.

\subsection{Quasi-pure approximation\label{quasi-pure}}
By now, a large number of experiments focusses on the (controlled) preparation
and evolution of entangled states. 
Any degree of mixing - \ie, the presence of classical correlations -- decreases
quantum 
correlations, and can lead to their complete destruction.
Therefore -- in particular in view of the many potential 
technical applications of non-classically correlated quantum states -- 
{\em pure} entangled states
are the experimentalist's desire:
hence it is crucial to 
screen the investigated
systems 
from the environment.

In general, perfect screening is impossible,
but experimental techniques are sufficiently advanced
\cite{wine95,hae,lew02} to preserve entanglement over 
appreciable periods of time.
Yet, very little is known on the precise temporal evolution of
such states even under weak but finite 
environment coupling,
one of the principal obstacles being 
the lack of computable entanglement measures for arbitrary states.

On the other hand, a general quantifier for entanglement, applicable to
arbittrary states, is not even needed in this context, 
since environmental influences can be assumed to be small under the given
experimental conditions.
Indeed the evolution of an initially pure into a mixed state 
occurs
on a rather long time scale, and
the experimentally interesting states 
are - though not exactly pure -
at least {\it quasi-pure},
\ie , they have one single eigenvalue $\ev_1$ that is much
larger than all the other ones.

In order to provide some efficient 
means to deal with this type of 
problems, 
we now derive an analytic approximation of concurrence for 
quasi-pure
states~\cite{floqp}. 
Indeed, we will find that this approximation 
also leads to a lower bound for arbitrary states.
This quasi-pure approximation will allow for an efficient quantitative
treatment of 
non-classical correlations that arise in most present day experiments.

The matrix ${\cal A}$ defined in Eq.~(\ref{elements-A}) contains the
matrix elements of $A$ evaluated with the subnormalised (see
\Eq{subnormalised}) eigenstates of $\varrho$.
Therefore, the elements of ${\cal A}$ are proportional to the 
eigenvalues $\ev_i$ of the considered state:
\be
{\cal A}_{jk}^{lm}\sim\sqrt{\bigl.\ev_j\ev_k\ev_l\ev_m\bigr.}\ .
\ee
Consequently, we can classify the elements of ${\cal A}$ according to
their relative magnitude determined by the eigenvalues $\ev_j$.
This classification will serve as a basis for the approximate evaluation
of concurrence in our subsequent treatment.

The above proportionality leads to a natural order of the elements of ${\cal
A}$, 
in terms of powers of square roots of the real eigenvalues
$\ev_i$ of $\varrho$,
which we assume to be labeled in decreasing order, \ie,
$\ev_1\gg\ev_2\ge\hdots\ge\ev_n$.
Hence, if we consider terms proportional to either one of the $\ev_j$,
with $j>1$, as perturbations of the dominant term
${\cal A}_{11}^{11}\sim\ev_1^2$,
we obtain the following classification:
\begin{itemize}
\item[-]the element ${\cal A}_{11}^{11}$ is lowest order,
\item[-]all elements with one index $j>1$, \ie
  ${\cal A}_{j1}^{11}$, ${\cal A}_{1j}^{11}$, ${\cal A}_{11}^{j1}$
  and ${\cal A}_{11}^{1j}$, are first order, and
\item[-]elements with two indices $j,k>1$, alike
  ${\cal A}_{jk}^{11}$ or ${\cal A}_{j1}^{k1}$, are second order.
\end{itemize}

However, this classification is not yet a sufficient basis for our approximation.
In fact, the element ${\cal A}_{11}^{11}$ is lowest order,
though still it could vanish.
This is the case if and only if the subnormalised eigenstate
$\ket{\psi_1}$ to the
largest eigenvalue $\ev_1$ of $\varrho$ is separable,
since $A_{11}^{11}$ is the square of the concurrence of $\ket{\psi_1}$
- see Eqs.~(\ref{concurrence-A}) and (\ref{elements-A}).
Therefore, as an additional requirement to quasi-purity we have to
impose that $\ket{\psi_1}$ is entangled.
Since the desired approximation is supposed to be applied to states
that occur in the experiments mentioned above,
this is not too stringent a restriction -
if an ideal experiment without any environment coupling led to a
pure state with non-negligible entanglement,
it is reasonable to assume that the eigenstate $\ket{\psi_1}$
associated with the largest eigenvalue $\ev_1$ of $\varrho$ is not separable
either. 

We wish to approximate ${\cal A}$ by a matrix product
\be
{\cal A}_{jk}^{lm}\simeq{\cal T}_{lm}^\ast{\cal T}_{jk}\ ,
\label{approx}
\ee
with a complex symmetric matrix ${\cal T}\in{\mathbbm C}^{n\times n}$.
Such a replacement allows for an analytic solution,
since the sum over $\alpha$ in
\Eq{concurrence-T} reduces to a single term, and the analytic
expression Eq.~(\ref{wotinfsol}) for the infimum, derived in
Section~\ref{concurrence-two-level}, can be employed.

Specifically for the lowest order term
${\cal A}_{11}^{11}$, \Eq{approx} 
yields ${\cal T}_{11}=\sqrt{{\cal A}_{11}^{11}}$,
up to an arbitrary phase which can be dropped.
Subsequently, 
evaluation of \Eq{approx} for the first order
elements
leads to
${\cal T}_{j1}={\cal A}_{j1}^{11}/\sqrt{{\cal A}_{11}^{11}}$, employing also
the expression for ${\cal T}_{11}$.
Finally, we still have the freedom to fix ${\cal T}_{jk}$, for $j,k\neq 1$.
For this purpose we use \Eq{approx} for ${\cal A}_{jk}^{11}$,
what leads to
${\cal T}_{jk}={\cal A}_{jk}^{11}/\sqrt{{\cal A}_{11}^{11}}$,
such that all matrix elements of $\cal T$ are given by
\be
{\cal T}_{jk}=\frac{{\cal A}_{jk}^{11}}{\sqrt{{\cal A}_{11}^{11}}}\ .
\ee
With this choice of $\cal T$, \Eq{approx} is exact at lowest and first
order, and - in addition - the second order elements
${\cal A}_{jk}^{11}$
are taken into account correctly.
Note that, using only one single matrix ${\cal T}$, it is not possible
to describe accurately {\em all} second order elements, such as ${\cal
A}_{j1}^{k1}$. 
All terms of third and fourth order are dropped in the present approximation.

Since we were assuming $\ket{\psi_1}$ to be entangled, \ie ,
${\cal A}_{11}^{11}$ finite by virtue of \Eq{concurrence-A},
${\cal T}$ is well defined.
Approximating ${\cal A}$ in terms of this matrix ${\cal T}$,
\Eq{concurrence-T} can be approximated as
\be
c(\varrho)\simeq c_{\rm{qp}}(\varrho)=\inf_V\sum_i
\left|\left[V{\cal T}V^T\right]_{ii}\right|\ ,
\label{cqp}
\ee
following the same steps as in the derivation of
\Eq{two-level-concurrence}. As formulated in \Eq{infimum}, the infimum can be
expressed in terms 
of the decreasingly ordered singular values $\sv_i$ of ${\cal T}$,
\be
c_{\rm{qp}}(\varrho)=
\max\left\{\sv_1-\sum_{i>1}\sv_i,\ 0\right\}\ .
\ee
A priori, it is not clear up to which degree of mixing the quasi-pure
approximation provides reliable estimates.
The fact that not all second order elements are taken into account may
somewhat reduce our confidence in a wide range of applicability for this
approximation. 
However, we will see in Sections~\ref{secrand}, \ref{dynmultpart}, and \ref{realist} that 
the above ansatz provides very good results for many states,
even for those with a substantial degree of mixing.

Indeed, the quasi-pure approximation is not only an approximation, but even a
lower bound of concurrence:
Given the decomposition
${\cal A}_{jk}^{lm}=\sum_{\alpha}(T_{lm}^\alpha)^\ast T_{jk}^\alpha$,
${\cal T}$ is easily expressed as
\be
{\cal T}=\sum_\alpha z_\alpha T^\alpha\ ,
\hspace{.5cm}\mbox{with}\hspace{.5cm}
z_\alpha=\frac{\left(T_{11}^\alpha\right)^\ast}
{\sqrt{\sum_\beta|T_{11}^\beta|^2}}\ .
\ee
Thus, ${\cal T}$ is indeed a valid symmetric matrix to provide a lower
bound as formulated in \Eq{lowb}.

\subsection{Lower bounds of states with positive partial
  transpose\label{seccompare}} 
In the preceding section we collected a set of operational tools
to access the non-classical correlations of arbitrary mixed quantum
states of finite dimension - characterised by their concurrence.
Our formal treatment spans the entire range from an approximation-free
description for numerical implementation,
over lower bounds that can be tightened numerically, to an easily tractable,
purely algebraic estimate of the degree of entanglement of a quasi-pure
quantum state which is typically dealt with in experiments. 
However, our approach is largely based on physical intuition,
and so far we cannot come up with mathematically precise error bounds.
Often, however, the latter are available in full mathematical 
rigour only under rather restrictive assumptions - while we are
seeking for robust quantities which, beyond formal consistency, can
cope with requirements which stem from real-world experiments.

Now we will test our toolbox under realistic conditions, and we
start out with the detection of nonseparable states with positive partial
transpose (ppt). In the next chapter, we will then use our approach to monitor
the time evolution of entanglement, under various scenarios. 

\subsubsection{Some exemplary ppt states\label{secppt}}

One of the main requirements imposed on an entanglement measure is
that it be able to distinguish entangled states from separable ones.
Whereas a large class of entangled states is detected by the ppt criterion
defined in \Eq{ppt},
no operational criterion is known so far that can detect all states
with positive partial transpose.
In general it 
tends to be rather demanding to decide whether such states
are entangled or not.
Therefore - albeit our bound, \Eq{lowb}, is capable of more than just checking
separability - we use it, as a first test of its pertinence, as a
separability criterion for some families of entangled states with
positive partial transpose~\cite{hor98,hor99,hor97}.
\begin{figure}
\epsfig{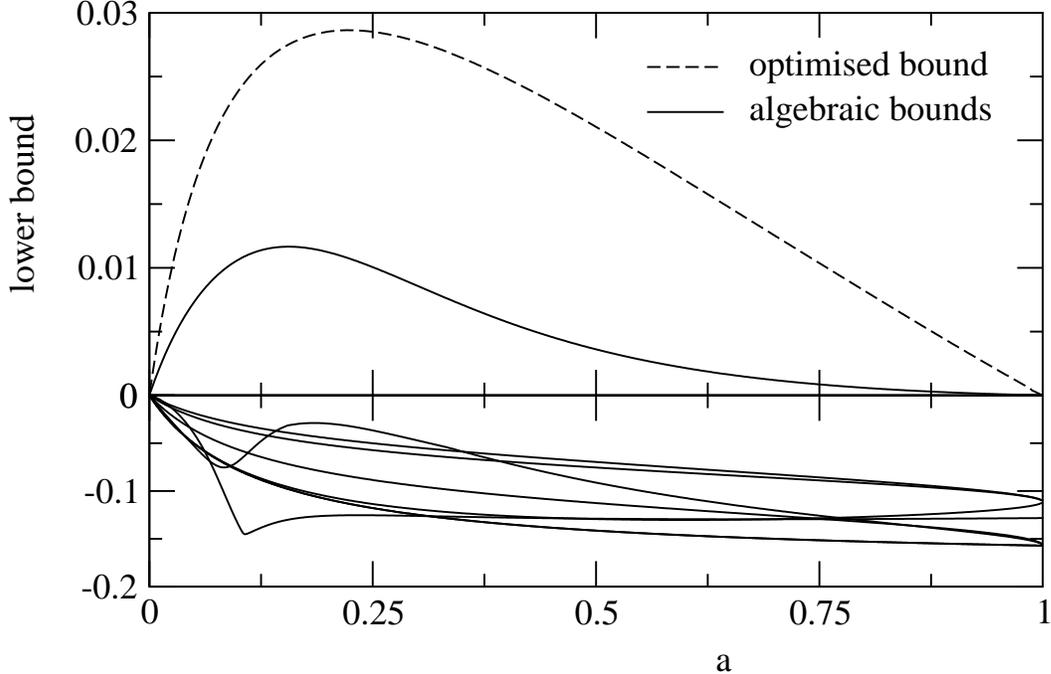}
\caption{
  Numerically optimised lower bound (dashed line,
  Eqs.~(\ref{bound}) and~(\ref{freeparbound})) of   
  the concurrence of the family of bipartite spin-1 states $\varrho_a$
 ~\cite{hor97} as defined in \Eq{hor33},
  together with the algebraic bounds (solid lines) obtained from Eq.~(\ref{bound})
  using~(\ref{T-pab}). Both the optimised bound and the largest
  algebraic bound are positive,
  such that the state is detected as entangled in
  the entire parameter range $a=[0,1]$.
  All other algebraic lower bounds are negative.
  Note that the scale of concurrence (y-axis) is different for positive and negative bounds.
}
\label{conchorlb}
\end{figure}

The first class of states describes a bipartite spin-$1$ system.
The state $\varrho_a$ acting on
${\mathbbm C}^3\otimes{\mathbbm C}^3$
is given~\cite{hor98,hor97}, for $a\in[0,1]$, as
\be
\varrho_a=\frac{1}{1+8a}\left[
{\renewcommand{\arraystretch}{0.8}\ba{ccccccccc}
a & 0 & 0 & 0 & a & 0 & 0 & 0 & a \\
0 & a & 0 & 0 & 0 & 0 & 0 & 0 & 0 \\
0 & 0 & a & 0 & 0 & 0 & 0 & 0 & 0 \\
0 & 0 & 0 & a & 0 & 0 & 0 & 0 & 0 \\
a & 0 & 0 & 0 & a & 0 & 0 & 0 & a \\
0 & 0 & 0 & 0 & 0 & a & 0 & 0 & 0 \\
0 & 0 & 0 & 0 & 0 & 0 & \beta & 0 & \gamma \\
0 & 0 & 0 & 0 & 0 & 0 & 0 & a & 0 \\
a & 0 & 0 & 0 & a & 0 & \gamma & 0 & \beta
\ea}\right],
\hspace{.5cm}\mbox{with}\hspace{.5cm}
\left\{
\ba{l}
\ds\beta=\frac{1+a}{2}\ ,\\
\left.\right.\\
\ds \gamma=\frac{\sqrt{1-a^2}}{2}\ ,
\ea
\right.
\label{hor33}
\ee
and has a positive partial transpose as defined in \Eq{ppt}, in
the entire range of $a$.
The algebraic lower bounds defined by Eqs.~(\ref{bound}) and~(\ref{T-pab}) 
are plotted in Fig.~\ref{conchorlb} as solid lines.
One of them is positive for all values of the parameter $a$,
and the non-separability of $\varrho_a$ is therefore detected by a
{\em purely algebraic criterion}.
All other algebraic bounds are negative, and therefore do not provide
any information on their own.

The dashed line in Fig.~\ref{conchorlb} shows the lower bound that is
numerically optimised over the $Z_\alpha$ in \Eq{lowb},
using a {\it downhill simplex method}~\cite{numrep}.
It is significantly larger than the positive algebraic bound,
and shows a qualitatively different behaviour for large $a$,
where its first derivative is finite, while that of the largest
algebraic bound vanishes for $a=1$.\\
\begin{figure}
\psfrag{b_1}{$b_1$}
\psfrag{b_opt}{$b_{opt}$}
\epsfig{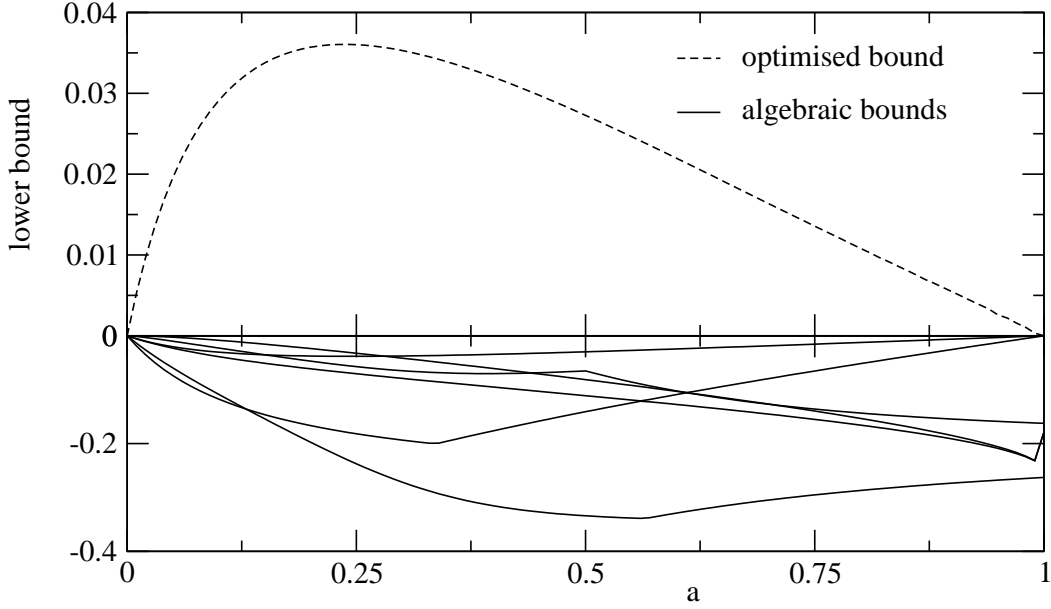}
\caption{
  Algebraic lower bounds  of the concurrence (see Eqs.~(\ref{bound})
  and~(\ref{T-pab}), solid lines) of a 
  family of $4\times 2$ states~\cite{hor97} defined by \Eq{hor24},
  plotted as a function of the parameter $a$.
  Although none of these bounds is positive,
  the optimised bound (dashed line), \Eq{bound}, {\em is} positive.
  Thus the state is detected as entangled in the entire parameter
  range of $a$.
}
\label{conchor42}
\end{figure}
A second class of states $\varrho_a$~\cite{hor97} acts on
${\mathbbm C}^4\otimes{\mathbbm C}^2$,
and once again has a positive partial transpose, \Eq{ppt}, for $a\in[0,1]$:
\be
\varrho_a=\frac{1}{1+7a}\left[
{\renewcommand{\arraystretch}{0.8}\ba{ccccccccc}
a & 0 & 0 & 0 & 0 & a & 0 & 0 \\
0 & a & 0 & 0 & 0 & 0 & a & 0 \\
0 & 0 & a & 0 & 0 & 0 & 0 & a \\
0 & 0 & 0 & a & 0 & 0 & 0 & 0 \\
0 & 0 & 0 & 0 & \beta & 0 & 0 & \gamma \\
a & 0 & 0 & 0 & 0 & a & 0 & 0 \\
0 & a & 0 & 0 & 0 & 0 & a & 0 \\
0 & 0 & a & 0 & \gamma & 0 & 0 & \beta
\ea}\right],
\hspace{.5cm}\mbox{with}\hspace{.5cm}
\left\{
\ba{l}
\ds\beta=\frac{1+a}{2}\ ,\\
\left.\right.\\
\ds \gamma=\frac{\sqrt{1-a^2}}{2}\ .
\ea
\right.
\label{hor24}
\ee
Figure~\ref{conchor42} shows the algebraic lower bounds obtained as
discussed in Section~\ref{algebraic-bound}, which are all negative.
Thus, none of them detects $\varrho_a$ as entangled.
Though, due to the degeneracy of this particular state, there is a
degeneracy in the eigenvalues of ${\cal A}$, such that the matrices
$T^\alpha$, and, consequently, also the algebraic lower bounds, are
not uniquely determined.
Neither did we 
find any matrices $T^\alpha$ in the degenerate subspaces
that provide positive lower bounds.
Yet, the numerically optimised lower bound - also shown in
Fig.~\ref{conchor42} -
{\em is} positive in the entire parameter range.
Hence, also this state is detected as entangled by our lower bound,
\Eq{bound}.\\ 

\begin{figure}
\epsfig{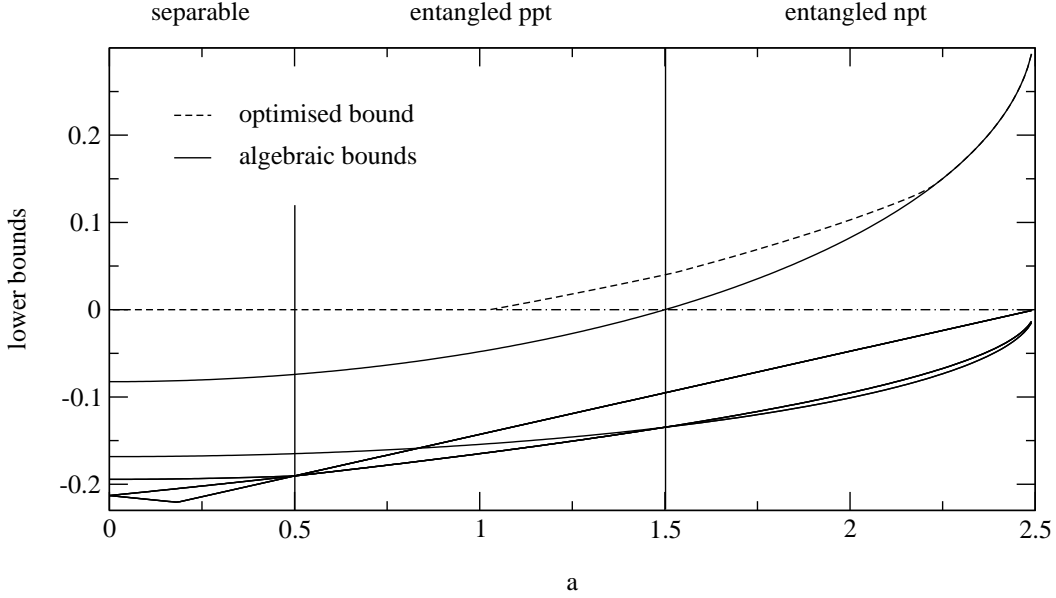}
\caption{
  Algebraic lower bounds (see Eqs.~(\ref{bound})
  and~(\ref{T-pab}), solid lines)
 of a family of $3\times 3$ states 
 ~\cite{hor99} defined in \Eq{horror}, as a function of the parameter $a$.
  For $a\le 1/2$, $\varrho_a$ is separable, and for $a>1/2$ it is
  entangled~\cite{hor99}.
  It has positive partial transpose for $a\le 3/2$, and non-positive
  partial transpose for $a>3/2$.
  The dashed line shows the numerically optimised lower bound, \Eq{bound}. 
  The state $\varrho_a$ is detected as entangled by the algebraic
  bound, exactly in that parameter range where it has non-positive
  partial transpose.
  It is detected as entangled by the optimised bound in approximately
  half the parameter range with positive partial transpose,
  what, however, might be due to a failure of our numerical
  optimisation routine.
  }
\label{conchorror2}
\end{figure}
A third class of states $\varrho_a$~\cite{hor99}, acting on
${\mathbbm C}^3\otimes{\mathbbm C}^3$,
is defined for $a\in[-5/2,5/2]$,
\be
\varrho_a=\frac{1}{21}\left[
{\renewcommand{\arraystretch}{0.8}\ba{ccccccccc}
2 & 0 & 0 & 0 & 2 & 0 & 0 & 0 & 2 \\
0 & \beta_- & 0 & 0 & 0 & 0 & 0 & 0 & 0 \\
0 & 0 & \beta_+ & 0 & 0 & 0 & 0 & 0 & 0 \\
0 & 0 & 0 & \beta_+ & 0 & 0 & 0 & 0 & 0 \\
2 & 0 & 0 & 0 & 2 & 0 & 0 & 0 & 2 \\
0 & 0 & 0 & 0 & 0 & \beta_- & 0 & 0 & 0 \\
0 & 0 & 0 & 0 & 0 & 0 & \beta_- & 0 & 0 \\
0 & 0 & 0 & 0 & 0 & 0 & 0 & \beta_+ & 0 \\
2 & 0 & 0 & 0 & 2 & 0 & 0 & 0 & 2 
\ea}\right],
\hspace{.5cm}\mbox{with}\hspace{.5cm}
\beta_\pm=\frac52\pm a\ .
\label{horror}
\ee
Since replacing $a$ by $-a$ is equivalent to exchanging the subsystems,
we will discuss this state only for $a\in[0,5/2]$.
The state $\varrho_a$ has a non-positive partial transpose for
$a\in[3/2,5/2]$,
is entangled with positive partial transpose for
$a\in[1/2,3/2]$ and is separable for $a\in[0,1/2]$~\cite{hor99}.
As depicted in Fig.~\ref{conchorror2}, $\varrho_a$ is detected as
entangled in its domain of negative partial transpose already by the
best algebraic lower bound.
In the regime where $\varrho_a$ has positive partial transpose all 
algebraic bounds are negative, such that the optimised lower bound is
required for distinguishing $\varrho_a$ from separable states.
However, even the optimised bound does not detect $\varrho_a$ in the
entire interval $a\in[1/2,3/2]$.
For $a\lesssim 1.02$, the lower bound seems to fail as a sufficient
separability 
criterion.
At present, we have no conclusive evidence from our numerical
optimisationroutine to decide whether the bound itself is not good enough, or
whether the numerically found maximum is not the global one.

The above exemplary $ppt$ states show that our lower bound, \Eq{bound},
is capable to detect families of 
entangled states which are not recognised by the
$ppt$ criterion.
For some states it is even not necessary to evaluate the optimised
bound, since already one of the algebraic bounds, introduced in
Section~\ref{algebraic-bound}, is positive.
Moreover, also the quasi-pure approximation is positive for some $ppt$
states \cite{floqp}.
However, there are also states with only negative algebraic bounds and
negative quasi-pure approximation, though positive optimised bound.
Our last example above showed a case of entangled states that we have
so far been unable to detect for a small subset of parameters,
though it remains hitherto undecided whether this is a failure of our
numerical optimisation routine or of our lower bound, \Eq{bound}, itself.
It is obvious from the different behaviour of optimised and algebraic bounds at
the border line between {\em ppt} and non-{\em ppt} regions in
Fig.~\ref{conchorror2}, that some more profound algebraic signatures remain to
be uncovered.

\section{Dynamics of entanglement under environment coupling \label{secdyn}}
Arguably the central motivation for deriving efficiently evaluable measures of
the entanglement of mixed states is the ubiquity of the latter in any real
physical setting. If we consider entanglement as the central resource of most
types of quantum information processing, then the experimentally most relevant
question is that of the lifetime of entanglement under the environment-induced
mixing. This is the subject of the present, concluding section of this
review. 

In a first subsection, we will test the mixed state entanglement estimates
derived in the previous sections, such as to demonstrate their versatility to
describe generic time evolutions under environment coupling. Here, the time
evolution both of the system and of the bath will be generated by random
Hamiltonians, without any specific physical realization in mind. 

Subsequently, we will specialise to particular, experimentally relevant (since
realized) cases, and specifically focus on the scaling properties of the
typical time scales which determine the time evolution of bipartite and
multipartite entanglement.    

\subsection{Random time evolution of higher-dimensional bipartite
  systems \label{secrand}} 
Let us first have a closer look at the performance of the various entanglement
estimates derived above, for a generic time evolution under environment
coupling. For that purpose,
we consider a bipartite system
and a third system serving as environment.
The bipartite system is initially prepared in a maximally entangled
pure state
\begin{equation}
\ket{\varPhi_0}=\frac{1}{\sqrt\dss}\sum_{i=1}^\dss\ket{ii}\ ,
\label{maxent}
\end{equation}
\ie , it is not entangled with the environment.
We then evolve the total system under a unitary dynamics generated by the
Hamiltonian
\be
H=H_{\rm se}+H_{\rm s}\otimes{\mathbbm1}_{\rm e} + {\mathbbm1}_{\rm
  s}\otimes H_{\rm e},
\label{randtimeU}
\ee
with a randomly chosen hermitian matrix $H_{\rm se}$ acting
on system and environment,
while a second randomly chosen hermitian matrix $H_{\rm s}$ only acts on the
(bipartite) system, but not on the environment. All elements of $H_{\rm se}$
and $H_{\rm s}$ are determined independently under the constraint of
hermiticity. Any real entry is obtained as $\sin(r)$, with a random integer
$r$, where the a priori probability is the same for any integer $0\le
r<10^{15}$~\cite{cbuch}. Furthermore, we neglect the free evolution of the
environment by setting $H_{\rm e}={\mathbbm1}_{\rm e}$.

Tracing out the environment after a finte interaction time $t$ leads to a mixed state of the bipartite system.
$H_{\rm s}$
describes the
interaction between the bipartite system's components
and does not induce any
mixing,
though it can change
the degree of entanglement.
We further scale the Hamiltonians $H_{\rm se}$ and $H_{\rm s}$ by the real coupling constants $\alpha_{\rm
  se}$ and $\alpha_{\rm s}$, which fix the relative time scales of unitary and incoherent
system dynamics.
$\alpha_{\rm se}$
determines the strength of the
system-environment interaction, and therefore
the mixing rate of the system,
whereas $\alpha_{\rm s}$ specifies the time-scale of the unitary system
evolution, causing a reversible decrease and increase of
bipartite
entanglement therein.

\begin{figure}
\epsfig{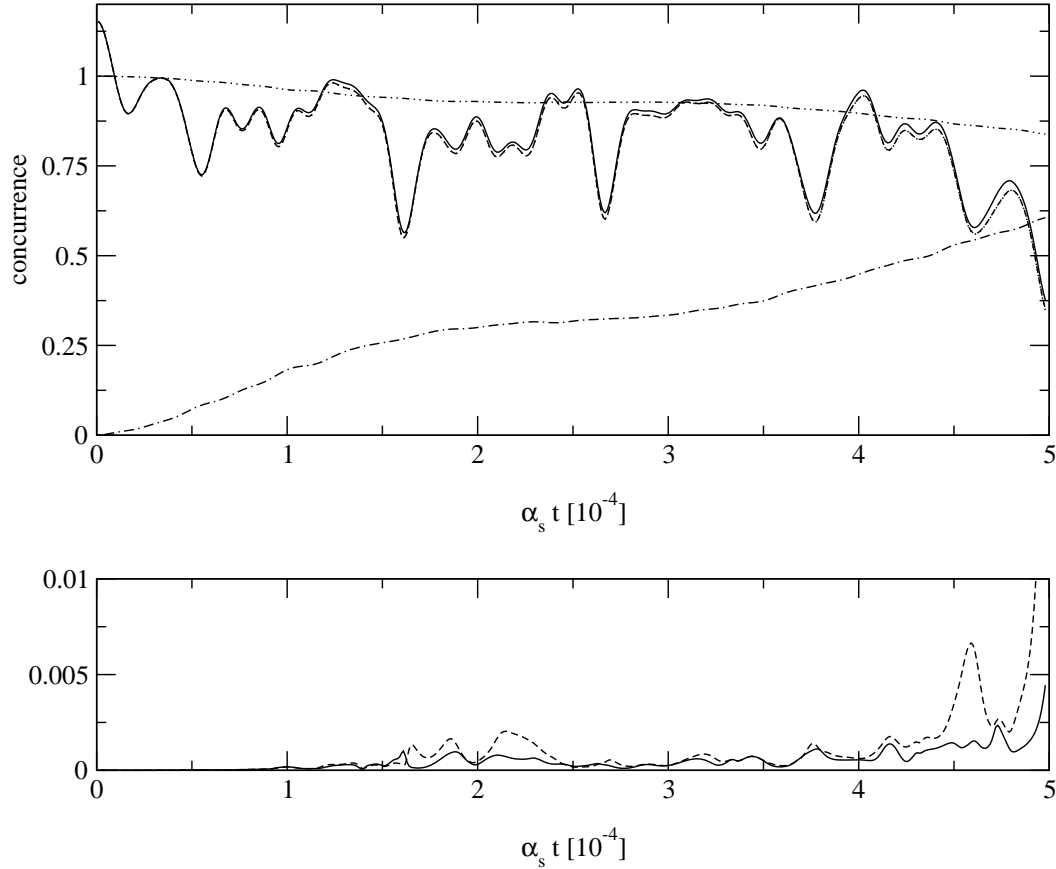}
\caption{
  Top panel: Upper bound (\Eq{upbound}, solid line) and numerically
  optimised lower bound (Eqs.~(\ref{bound}) and~(\ref{freeparbound}), dashed line) of concurrence of a
  $3\times 3$ state, vs. scaled time $\alpha_{\rm s}t$.
  The system
  is initially prepared in a maximally entangled state of type (\ref{maxent}),
  with concurrence $c=2/\sqrt3$ (see Eq.~(\ref{maxconcI})).
  A random
  time evolution according to (\ref{randtimeU}), with $\alpha_{\rm se}/\alpha_{\rm s}=10^{-2}$, leads to a finite degree of mixing measured by
  the system state's von Neumann entropy (dash-dotted line). The largest
  eigenvalue of $\varrho$ evolves along the dash-double-dotted line.
In the lower panel, the solid and the dashed line represent the difference
  between the optimized lower bound of concurrence, Eqs.~(\ref{bound})
  and~(\ref{freeparbound}),
  and the quasi-pure approximation, Eq.~(\ref{cqp}), or the best algebraic
  lower bound, Eqs.~(\ref{bound}) and~(\ref{T-pab}),
  respectively. As also visible from the upper panel, all our lower
  bounds provide excellent estimates of the actual value of concurrence over
  the entire evolution period, almost undistinguishable from the upper bound,
  Eq.~(\ref{upbound}) (full line in the upper panel).
}
\label{fig334}
\end{figure}

With these premises, we can now monitor the time evolution of the concurrence
under the above random dynamics,
as a function of the scaled time $\alpha_{\rm s}t$,
for different values of the ratio $\alpha_{\rm se}/\alpha_{\rm s}$.
The degree of mixing of $\varrho$ will be characterised by the von Neumann entropy
$S=-\tr\varrho\ln\varrho$ of the system state.
In addition, we also
follow the time evolution of the largest eigenvalue of $\varrho$,
since this was the central pillar in our derivation of the quasi-pure approximation (Section~\ref{quasi-pure}).

\begin{figure}
  \epsfig{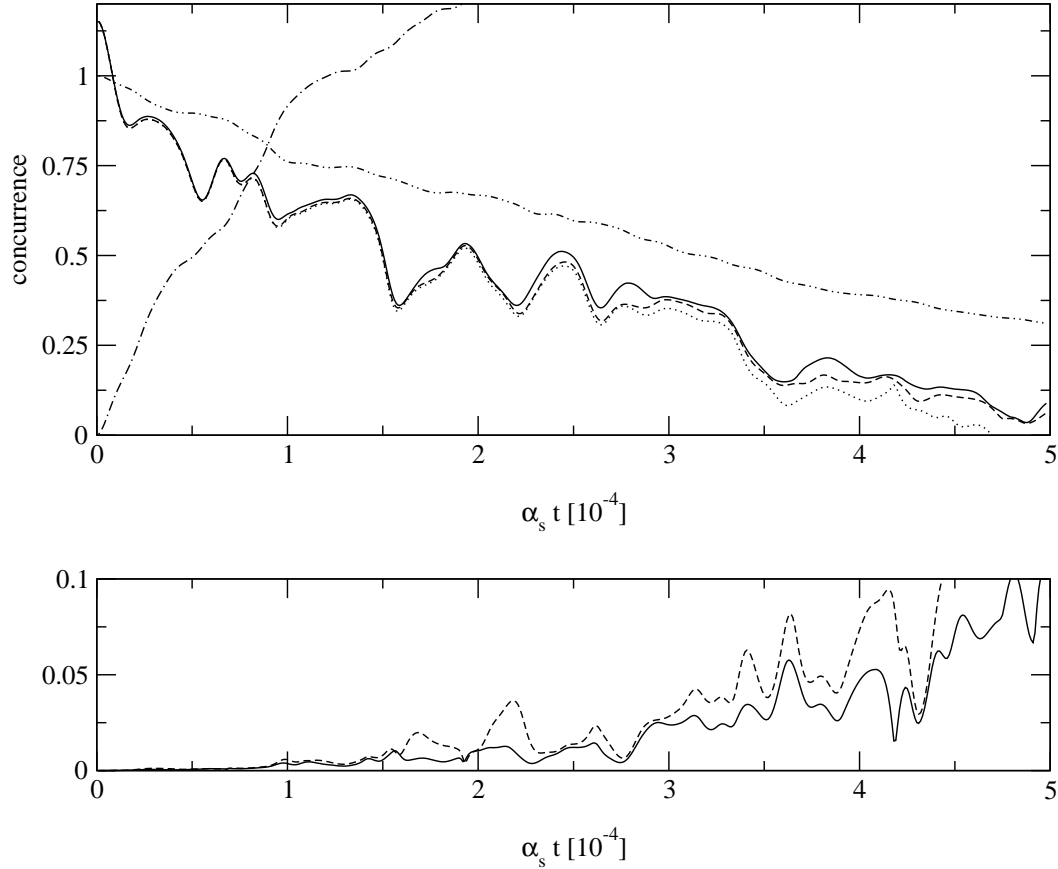}
\caption{
Same as Fig.~\ref{fig334}, with a larger environment coupling constant
$\alpha_{\rm se}/\alpha_{\rm s}=5\times 10^{-2}$.}
\label{fig335}
\end{figure}

The upper panels in Figs.~\ref{fig334} and~\ref{fig335} show the time
evolution of upper (\Eq{upbound},
solid line) and lower (Eqs.~(\ref{bound}) and~(\ref{freeparbound}), dashed line) bounds of
concurrence, together with concurrence in quasi-pure approximation (\Eq{cqp},
dotted line), for two different random time evolutions of the same maximally
entangled $3\times 3$ initial state. The initial value of $c(\varrho)$ follows
immediately from Eq.~(\ref{maxconcI}).
The interaction Hamiltonians $H_{\rm se}$ and $H_{\rm s}$ are the same
in both figures,
but the coupling constant $\alpha_{\rm se}$ in Fig.~\ref{fig335} is by
a factor five larger as compared to its value $\alpha_{\rm se}/\alpha_{\rm s}=10^{-2}$ in
Fig.~\ref{fig334}.
Consequently, mixing increases slower in Fig.~\ref{fig334} than in Fig.~\ref{fig335}.
The degree of mixing is characterised by both the largest eigenvalue
of $\varrho$, depicted as a dash-double-dotted line, and by its von Neumann
entropy (dash-dotted).

\begin{figure}
  \epsfig{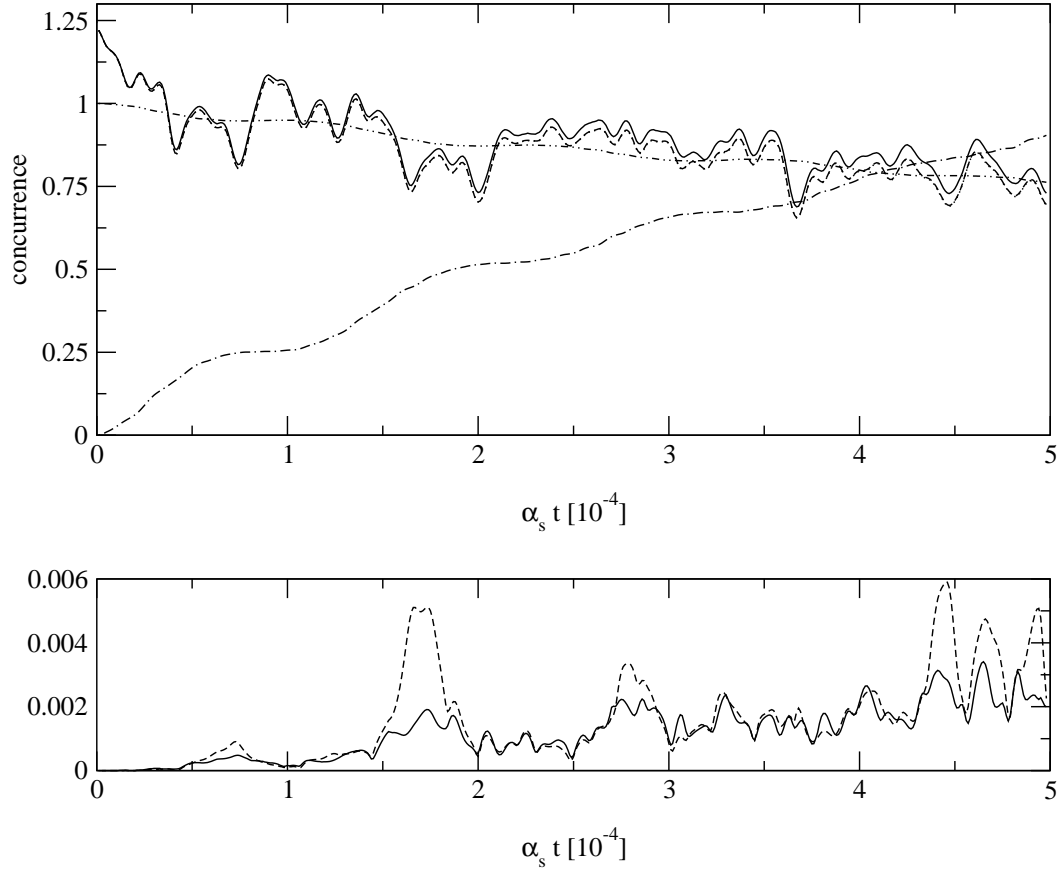}
\caption{
Same as Fig.~\ref{fig334}, for a $4\times 4$ system initially prepared in the
maximally entangled state, Eq.~(\ref{maxent}), and $\alpha_{\rm
  se}/\alpha_{\rm s}=10^{-2}$.
}
\label{fig443}
\end{figure}

For both values of $\alpha_{\rm se}/\alpha_{\rm s}$, the quasi-pure approximation and the
optimized lower bound remain almost indistinguishable, over the entire
system-environment interaction time.
The difference of both
quantities is plotted in the
lower panels of the figures, together with
the difference
between the numerically optimised lower bound, Eqs.~(\ref{bound}) and~(\ref{freeparbound}), and the
best algebraic bound, Eqs.~(\ref{bound}) and~(\ref{T-pab}).
Over almost the entire time interval displayed in Fig.~\ref{fig334},
and for short times in Fig.
\ref{fig335}, where mixing is not too large, the relative error of both
approximations is
about two orders of magnitude smaller than the actual value of
concurrence itself (note the different scales on the vertical axes).
Only states with rather small concurrence and a large degree of mixing
exhibit a significant difference between the optimized
lower bound and the best algebraic bound or
the quasi-pure approximation.

\begin{figure}
  \epsfig{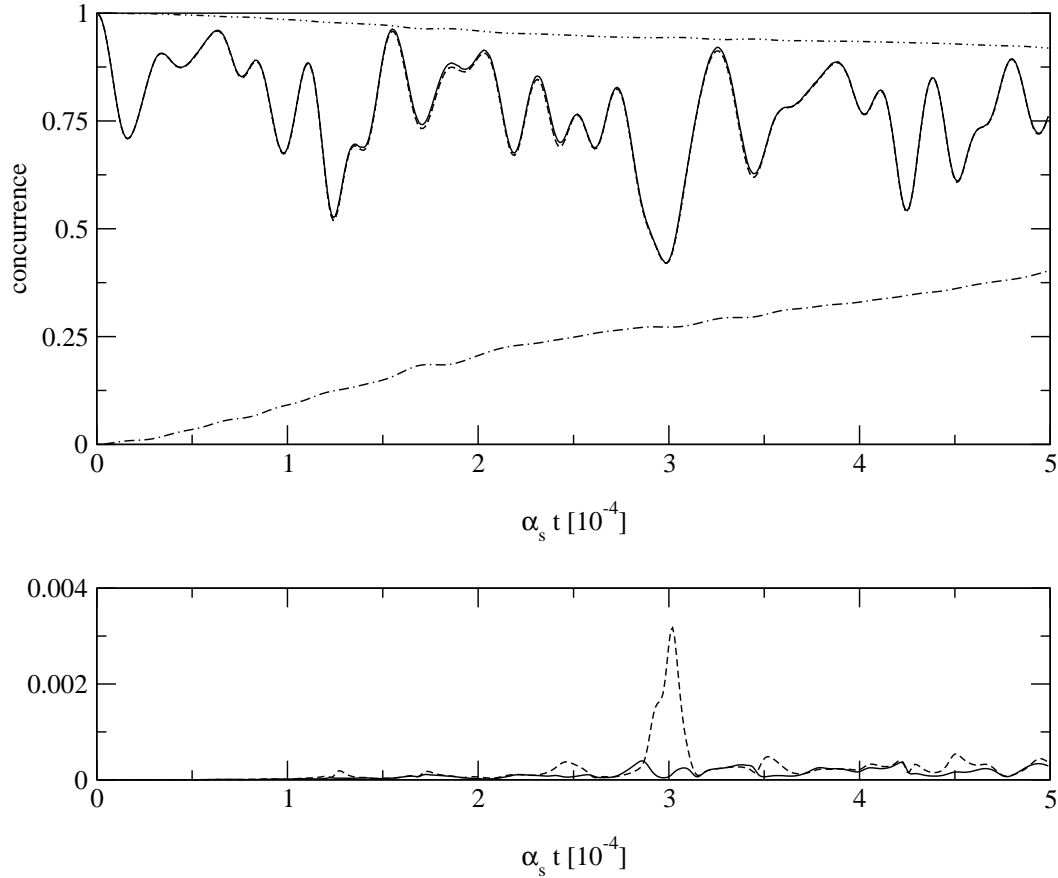}
\caption{
Same as in Figs.~\ref{fig334} and \ref{fig443}, for a $2\times 4$ system initially
prepared in the maximally entangled state, Eq.~(\ref{maxent}), and $\alpha_{\rm
  se}/\alpha_{\rm s}=10^{-2}$.
}
\label{fig241}
\end{figure}

Fig.~\ref{fig443} repeats the scenario of Fig.~\ref{fig334},
for a
$4\times 4$ system, with
environment coupling
constant $\alpha_{\rm se}/\alpha_{\rm s}=10^{-2}$.
The degree of mixing remains moderate during the entire time evolution,
as indicated by the von Neumann entropy
and the
largest eigenvalue of $\varrho$.
Once again, in this higher-dimensional system, upper,
Eq.~(\ref{upbound}), and optimised lower bound, Eq.~(\ref{bound}) and~(\ref{freeparbound}), provide an excellent
estimation of the actual value of concurrence. Equally so, also the best
algebraic lower bound, Eqs.~(\ref{bound}) and~(\ref{T-pab}), and the quasi-pure approximation perform
very well, over the entire time interval.
\begin{figure}
  \epsfig{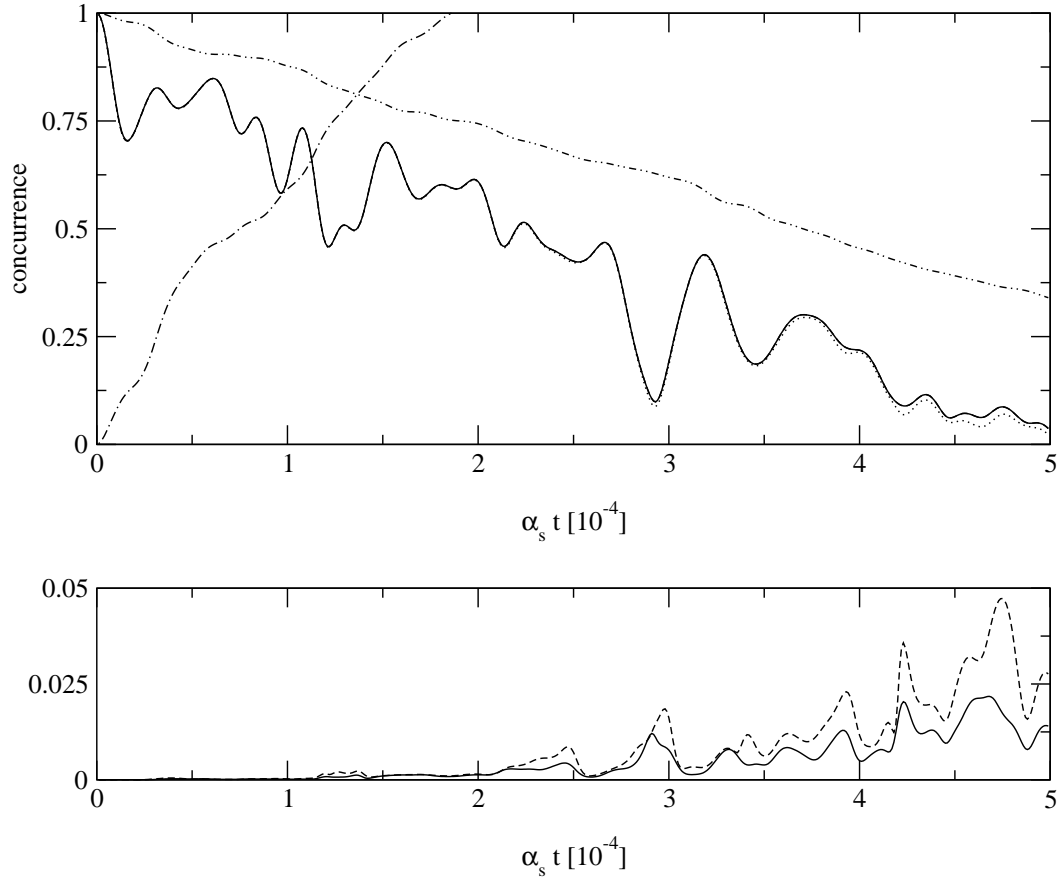}
\caption{
Same as Fig.~\ref{fig241}, but for a larger system-environment coupling
$\alpha_{\rm se}/\alpha_{\rm s}=5\times 10^{-2}$.
}
\label{fig242}
\end{figure}

Qualitatively the same observations are made for the time evolution of an
initially maximally entangled state, Eq.~(\ref{maxent}), of a $2\times 4$ system,
monitored in Figs.~\ref{fig241} and \ref{fig242}, for different
system-environment coupling strengths $\alpha_{\rm se}/\alpha_{\rm s}=10^{-2}$ and $5\times
10^{-2}$, respectively.
Only during a rather short time interval, when concurrence exhibits a dip at
$\alpha_{\rm s}t\simeq 3\times 10^{-4}$ in Fig.~\ref{fig242} does the
discrepancy between optimal lower bound and its estimates increase to
about one percent of the actual value of
concurrence, which, however, rather highlights the excellent reliability of
our estimates for all practical purposes.
In particular, this holds true for the quasi-pure approximation, albeit the
strongly mixed states occuring as time evolves in Fig.~\ref{fig242} hardly
satisfy the basic assumptions made for its derivation (see Section
\ref{quasi-pure}).

\subsection{Realistic scenarios of entanglement dynamics}
Recently, many groups were able to prepare entangled
states in a variety of physical systems and experimental setups, demonstrating
an impressive ability to manipulate and detect
them efficiently
\cite{wein00,hae,wine00,panPRL01,eiblPRL04,roos,rauschSCI00,zhaoNAT04,pashNAT03,berkleySCI04,roosPRL04}.
Particular
effort has been devoted to entangle multiple subsystems, not only to
investigate the possibility of fulfilling the fundamentally important 
scalability requirements for
quantum computation, but also to understand how far one can push a quantum
system towards the
macroscopic limit and still observe entanglement -  an intrinsic quantum
feature with no 
classical counterpart. 

However, a major obstacle for the controlled entanglement of more and 
more subsystems 
remains with the 
incapacity of achieving perfect screening 
of the
system from the environment. After some time, 
the unavoidable residual 
interaction with the reservoir induces mixing of the system state, and thus
the emergence 
of classical
correlations, at the expense of 
quantum entanglement.
Hence, we face the highly relevant task 
of understanding the sources of entanglement decay, what implies the 
identification of the associated time scales.

Although a general solution to this problem, for arbitrary system dynamics and
decoherence mechanisms is still out of reach, our technical machinery
developed in the previous sections allows to treat arguably all  situations
encountered in typical state of the art experiments, as we shall outline in
the sequel. We start with a short recollection of environment models adapted
for decoherence processes in a typical quantum optical context.

\subsubsection{Environment models \label{environ}}
We already anticipated in Section \ref{secrand} that
an open system can be described as a system $S$ which interacts with
an environment $E$, such that the dynamics of the total system $S+E$
is unitary, and governed by the von-Neumann equation
\begin{equation}
\frac{d{\varrho}_{se}}{dt}=-\frac{i}{\hbar}\left[H,{\varrho}_{se}\right],
\end{equation}
with the total Hamiltonian of the form~(\ref{randtimeU}).
However, we are only interested in the properties of the system $S$
itself.
An equation for the evolution of the (reduced) system density
matrix $\varrho$
is obtained by performing a trace over the
environmental degrees of freedom:
\begin{equation}
\frac{d{\varrho}}{dt}=-\frac{i}{\hbar}\tr_e\left[H,{\varrho}_{se}\right]\ .
\end{equation}
In Section~\ref{secrand} we considered $S$ as a bipartite system and
explicitely performed the total evolution given by $H$, obtaining the
reduced density matrix after tracing over environmental degrees of freedom. However, an equation only for the density matrix of the
system, after an interaction time $t$, can be obtained. Under the
assumption of complete positivity and Markovian dynamics,
it can be written in the
Lindblad form \cite{lind_76,kossak76,alicki_87}
\begin{equation}
\label{lindop}
\frac{d{\varrho}}{dt}=-\frac{i}{\hbar}\left [H_S,\varrho\right ] +{\cal L}  \varrho=
-\frac{i}{\hbar}\left [H_S,\varrho\right ] +
\sum_i\frac{\Gamma_i}{2}\left(2\,{d}_i\,{\varrho}\,{d}_i^\dagger -
{d}_i^\dagger\,{d}_i\,{\varrho} -
{\varrho}\,{d}_i^\dagger\,{d}_i\right)\, ,
\end{equation}
where operators ${d}_i$
describe the system-environment
coupling, with strength $\Gamma_i$.
The use of the Markovian
approximation is well justified in a large variety of quantum optical
experiments where entanglement has been produced, although one should
mention that non-Markovian effects can be important in the description of some
condensed-matter systems~\cite{loss02}.

Different situations may arise when a system is coupled to the
environment: energy can be exchanged and dissipation can take
place, noise can be added to the system, or
elastic processes can introduce loss of phase coherence without energy
transfer.
All these processes can be
described in terms of the above master equation by a suitable choice
of the operators ${d}_i$, which can be written, in the case of two-level
systems, in terms of the Pauli matrices.
For a two-level system interacting with a thermal bath,
for example, the non-unitary part of the master equation reads
\cite{mtqo}
\begin{equation}
\label{thermal}
{\cal L}\varrho = \frac{\Gamma (\bar n +1)}{2} \left(2\, \sigma_- \varrho \sigma_+ -
\sigma_+ \sigma_- \varrho - \varrho \sigma_+ \sigma_- \right) +\frac{\Gamma
  \bar n}{2}  \left(2\, \sigma_+ \varrho \sigma_- -
\sigma_- \sigma_+ \varrho - \varrho \sigma_- \sigma_+ \right) \, .
\end{equation}
In this equation, the first and the second term on the right hand side
describe, respectively, decay and
excitation processes, with rates which depend on the temperature, here
parametrised by $\bar n$,
the average thermal excitation of the
reservoir. In the limit of vanishing
temperature, ${\bar n}=0$,
only the
spontaneous decay term survives, leading to a purely dissipative
process,
\begin{equation}
\label{T0}
\frac{d{\varrho}}{dt}= \frac{\Gamma}{2} \left(2\, \sigma_- \varrho \sigma_+ -
\sigma_+ \sigma_- \varrho - \varrho \sigma_+ \sigma_- \right) \, ,
\end{equation}
which drives the system, asymptotically, to its ground state.

Another important limiting case of equation~(\ref{thermal}), which
describes noisy dynamics, is obtained for infinite temperature,
where $\bar n\rightarrow\infty$, and, simultaneously,
$\Gamma\rightarrow0$, so that $\Gamma \bar n \equiv
\tilde \Gamma$ remains constant:
\begin{equation}
\label{TI}
{\cal L}\varrho = \frac{\tilde \Gamma}{2} \left(2\, \sigma_- \varrho \sigma_+ -
\sigma_+ \sigma_- \varrho - \varrho \sigma_+ \sigma_- \right) +\frac{\tilde
  \Gamma}{2}  \left(2\, \sigma_+ \varrho \sigma_- -
\sigma_- \sigma_+ \varrho - \varrho \sigma_- \sigma_+ \right) \, .
\end{equation}
In this case, decay and excitation
occur at exactly the same rate, and
the noise induced by the transitions between the two levels brings the
system to a stationary, maximally mixed state.

A purely dephasing reservoir
is obtained by choosing $d_i = d = \sigma_+ \sigma_-$
in Eq.~(\ref{lindop}) leading to the master equation
\begin{equation}
\frac{d{\varrho}}{dt}= \frac{\Gamma}{2} \left(2\, \sigma_+ \sigma_-
  \varrho \sigma_+ \sigma_- - \sigma_+ \sigma_- \varrho - \varrho \sigma_+ \sigma_- \right) \, .
\label{Deph}
\end{equation}
In this case, there are no changes in the
populations of the ground and excited states, and
energy is conserved. Only the off-diagonal elements of the reduced
density matrix decay, leading to a loss of phase coherence.

In the context of entanglement, we have to describe how the
environment acts on the composite system. In Sections~\ref{dynbipart}
and \ref{dynmultpart} we
will assume that each subsystem interacts independently with the
environment, a well justified assumption whenever the particles
composing your system are sufficiently separated from each other and, therefore, no
collective environment effects must be taken into account. Hence, each
particle is subjected to a dynamics described by its own Lindblad
operator, supposed to be of the same form for all the components of
the system. Note that this assumption of mutually independent
environment couplings immediately implies that all initially entangled states will
asymptotically evolve into a separable state. We shall therefore focus on the precise timescales of this decay process.

In the final Section~\ref{realist}, we will consider an entanglement scheme~\cite{mol99} which
has been experimentally implemented using ion traps~\cite{wine00}, and in
which decoherence acts indirectly through the coupling of the collective
center-of-mass motion of the ions with the environment. Here, again,
environment interaction has a detrimental effect, which we shall study
in more detail focusing on the scaling of multipartite entanglement with the
system size (i.e., the number of entangled particles).

\subsubsection{Entanglement dynamics of bipartite two-level systems \label{dynbipart}} 
For bipartite two level systems, one can use the exact 
expression, Eq.~(\ref{wotinfsol}), for
concurrence, 
to
follow the time evolution of entanglement. While much simpler than the
multipartite case, the bipartite situation is useful for
developing some intuition about the entanglement decay under different
environment dynamics.

\begin{figure}
\includegraphics[width=14.0cm]{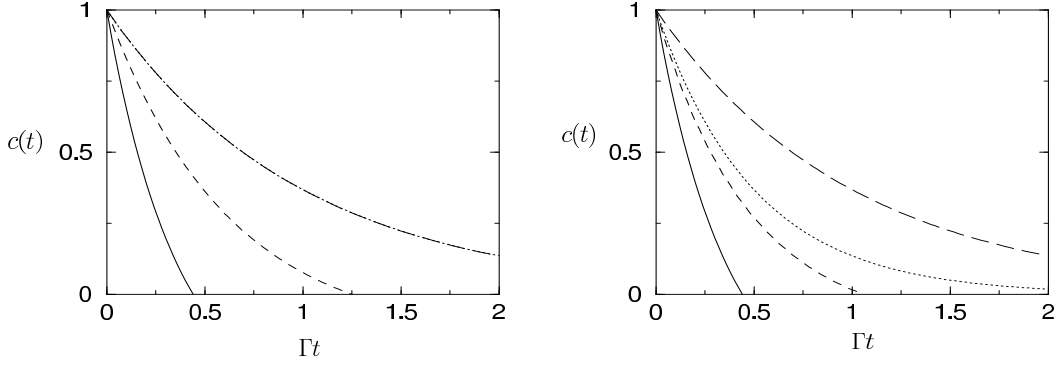}
\caption{Time dependence of the concurrence of a bipartite two-level system
initially prepared in Bell states $\vert \Psi^{\pm} \rangle$
  (left panel) or $\vert \Phi^{\pm} \rangle$ (right panel), under coupling to a zero
  temperature (dotted lines), dephasing
(long dashed lines), infinite (solid lines), and finite temperature
  ($\bar n=0.1$, dashed lines) environment (time evolution generated by
  Eqs.~(\ref{T0}), (\ref{Deph}), (\ref{TI}), and (\ref{thermal}),
  respectively. In the case of $\vert \Psi^{\pm} \rangle$ states, the 
  zero temperature and dephasing solutions coincide. In contrast, the zero
  temperature environment induces a twice as fast decay as compared
  to the dephasing environment, for the $\vert \Phi^{\pm} \rangle$ state. For all positive
  temperatures these initial states evolve into separable states in a
  finite time. 
Only zero temperature and 
dephasing environments induce separability 
only in the limit $t\rightarrow\infty$.} 
\label{concbipartfig}
\end{figure}

To start with, we 
consider an initial Bell state 
$\vert \Psi^{\pm}\rangle=\left(\vert 01 \rangle \pm \vert 10
  \rangle\right)/\sqrt{2}$. The solution of the master
equation for the different environment couplings discussed
previously is straightforward, and the time evolution
of 
concurrence, shown in Fig.~\ref{concbipartfig}, is given by
\begin{equation}
c(t)={\rm max}\left\{\frac{e^{-4\tilde \Gamma t}}{2}+e^{-2\tilde \Gamma
t}-\frac{1}{2},0\right\}\, ,  
\end{equation}
for the infinite temperature case, and by
\begin{equation}
c(t)=e^{-\Gamma t},
\end{equation}
for dephasing, Eq.~(\ref{Deph}), and zero temperature, Eq.~(\ref{T0}),
  environments. The situation is the
  same for the Bell state $\vert \Phi^{\pm}\rangle=\left(\vert 00 \rangle \pm \vert 11
  \rangle\right)/\sqrt{2}$ (right panel of Fig.~\ref{concbipartfig}),
  apart from the zero temperature case 
  where the concurrence decays as $c(t)=e^{-2\Gamma
  t}$. This accelerated (by a factor of two) decay of concurrence for
  the $\vert \Phi^{\pm} \rangle$ states as
compared to the $\vert \Psi^{\pm} \rangle$ states, under the influence of zero temperature
environment, can be 
understood from the time scales involved in the corresponding solution
for the density matrix: while for $\vert \Phi^{\pm} \rangle$ each term $\ket{01}$ and
$\ket{10}$ corresponds to a single particle decay, leading to a time
scale $e^{-\Gamma t}$, we have the
term $\ket{11}$ in $\vert \Psi^{\pm} \rangle$, such that both particles can undergo an
environment induced transition to the ground state, thus introducing a faster, $e^{-2 \Gamma
  t}$, decay.\footnote{Although intuitive and reasonable in this case, one
  must be carefull while using such kind of arguments to deduce the
  actual behaviour of concurrence from the time scales which appear in the
  coherences of $\varrho$. In general, there is no simple and obvious
  relation between these quantities.} General solutions
  for the finite temperature case can be given explicitely~\cite{arrc}, though are rather
  cumbersome and will not be
  presented here. However, one may expand these solutions and obtains, at
  first order in $t$, 
  \begin{equation}
    c(\Psi^{\pm},t)\simeq 1-\left(2\bar n+1 +2\sqrt{\bar n (\bar n
    +1)}\right)\Gamma t , 
  \end{equation}
and
  \begin{equation}
    c(\Phi^{\pm},t)\simeq 1-2(2\bar n+1) \Gamma t , 
  \end{equation}
for $\vert \Psi^{\pm} \rangle$ and $\vert \Phi^{\pm} \rangle$ states, respectively. These expressions not
  only reproduce the leading order terms of the zero and infinite
  temperature solutions in the appropriate limits, but also show the
  influence of the temperature on the short time behaviour of
  concurrence decay. 

It is equally simple to obtain the long-time asymptotics 
for the singular values of the matrix $\tau$ defined in
Section~\ref{secwotconc}, and, consequently, for the quantity that
enters expression (\ref{wotinfsol}) 
for concurrence. It reads, both for $\vert \Psi^{\pm} \rangle$ and $\vert \Phi^{\pm} \rangle$ states,
\begin{equation}
\lim_{t \to \infty} \left(\sv_1-\sum_{i=2}^4\ \sv_i\right) = -\frac{2\bar
      n(\bar n+1)}{(2\bar n+1)^2}\ . 
\end{equation}
This quantity is non-positive for all values of $\bar n$ and vanishes only
in the case of zero temperature. This means that 
the above initial
states evolve into separable states
within a finite time, 
for any $T>0$. 
In contrast, for zero temperature as well as for
dephasing environments does $c(t)$ only vanish in the limit
$t\rightarrow\infty$.\footnote{Note that the above is not true for general states. In
fact, some initial mixed states may reach separability on 
finite time scales also in the zero temperature
case~\protect\cite{eberly_2004}.}

\subsubsection{Entanglement dynamics of multipartite two-level systems
  \label{dynmultpart}} 
Let us now generalise our above investigation for larger arrays of 
two-level systems, with variable size $N$. These are objects often
encountered, e.g.,  in ion trap quantum computing schemes, and the robustness
of their entanglement properties with increasing size is at the very heart of
the fundamental scalability requirement on which hinges any future
technological application.

We have already prepared a comfortable tool for such generalisation -- the
multipartite concurrence $c_N$ defined in Eq.~(\ref{cN}), with the
particularly useful property 
\begin{eqnarray}
c_N(\ket{\varPhi}\otimes\ket{\varphi}) & = & c_{N-1}(\ket{\varPhi})\ , \\
{\rm for}\ \ket{\varPhi}\otimes\ket{\varphi} & \in & {\cal H_1}\otimes\ldots\otimes
{\cal H_N},\ {\rm and}\ \ket{\varphi}\in {\cal H_i},\ i\in\left \{
1,\ldots ,N\right \}\ .
\end{eqnarray}
This allows for the direct comparison of the entanglement properties of pure
and mixed states of multipartite qubit arrays of increasing size~\cite{floandre}, where the
mixed state concurrence is once again evaluated through the convex
roof Eq.~(\ref{mixed-concurrence}), with the appropriate definition of the operator $A$
in Eq.~(\ref{defA}).

Since the reliability of the lower bounds derived in Section~\ref{seclowb} was
demonstrated extensively in Section~\ref{secrand} above, we will
actually use the quasi-pure approximation Eq.~(\ref{cqp}) in the following, minimising the numerical effort very considerably, notably for large system
sizes $N$ (notwithstanding, some additional tests were performed at randomly
chosen instances, to convince ourselves of the quality of the quasi-pure
approximation with 
respect to the optimal upper and lower bounds, Eqs.~(\ref{upbound}) and
(\ref{bound}), 
respectively. 
 
\begin{figure}
\begin{center}
\includegraphics[width=8.0cm]{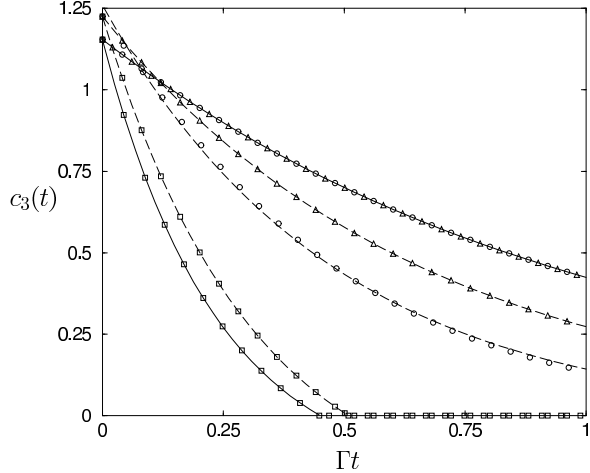}
\end{center}
\caption{Time dependence of the multipartite concurrence $c_N(t)$,
Eq.~(\ref{cN}), for 
$N=3$ particle systems initially prepared in a GHZ (dashed lines) or W
(solid lines) state, under the influence of 
a zero temperature (circles), dephasing
(triangles), and infinite temperature (squares) environments. In all
cases, the numerical results are very well fitted by an exponential decay
(solid and dashed lines).} 
\label{decay3part}
\end{figure}

We shall consider two types of initial states, somewhat similar to the singlet and
triplet states in Section~\ref{dynbipart} above: 
the GHZ state
\begin{equation}
\vert \Psi_{N}\rangle_{\rm GHZ}=\frac{1}{\sqrt{2}}\left ( \vert 00 \hdots 0
\rangle + \vert 11 \hdots 1 \rangle \right )\ , 
\label{ghzstate}
\end{equation}
and the W state, 
\begin{equation}
\vert \Psi_{N}\rangle_{W}=\frac{1}{\sqrt{N}}\left ( \vert 00 \hdots 01
\rangle + \vert 00 \hdots 10 \rangle + \hdots +\vert 10 \hdots 00
\rangle \right ).
\label{wstate}
\end{equation}
These were recently 
produced
\cite{wine00,roos} in the lab, for $N=3,4$, and it is now within
experimental reach to monitor the time dependence of their degree of
entanglement, by means of quantum state
tomography~\cite{roosPRL04}. 
Figure~\ref{decay3part} shows the decay of the concurrence
for tri-partite GHZ 
and W 
states under
the influence of zero temperature (circles), dephasing (triangles) and
infinite temperature (squares) environments as a function of time. The
symbols represent the result 
of the quasi-pure
approximation, while the lines indicate the best fit to 
a mono-exponential
decay $c_N(t)=Ae^{-\gamma t}+B$. 
As in the
bipartite case, the zero temperature and
dephasing environment lead to separability only in the limit $t \to \infty$,
for these initial states.~\footnote{Note that, as in the bipartite
  case, this is not true for general states. The state
  $\left(\ket{000} + \ket{011} +\ket{101} +\ket{110}\right)/2$, for
  example, reaches separability, according to quasi-pure calculations,
  on a finite time scale also in the zero temperature case.}

\begin{figure}
\begin{center}
\includegraphics[width=7.5cm]{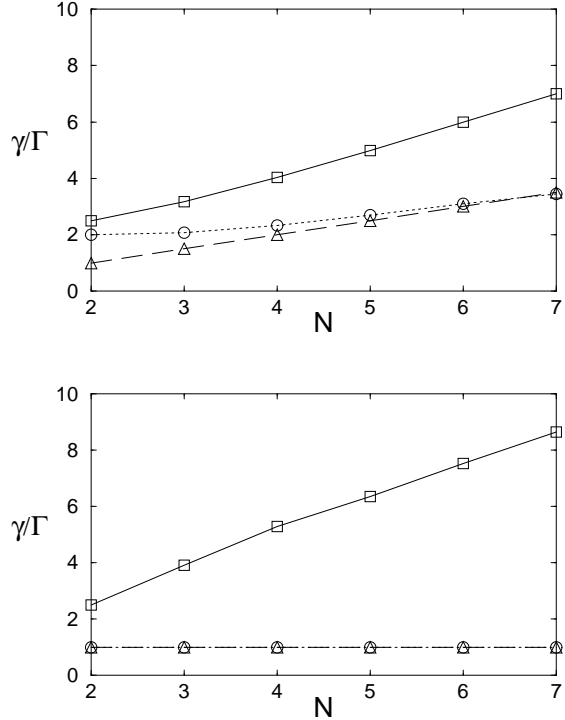}
\end{center}
\caption{
Concurrence decay rates $\gamma$ (in units of the reservoir rate $\Gamma$) for
GHZ (top) and W (bottom) states, as a function of the system size (i.e., 
the particle number)
$N$. The different 
environment models are represented by circles 
(zero
temperature), squares 
(infinite temperature), and
triangles 
(dephasing). The connecting lines are drawn to guide the eye. 
Whilst for GHZ states the
decay rates increase roughly linearly with $N$ (for sufficiently large $N$), 
independently of the specific
environment,
the W states exhibit increasing decay with system size 
only for the infinite temperature environment. Remarkably, the decay rate of
the W states is size-independent for dephasing and zero temperature environments!}
\label{gammas}
\end{figure}

To assess the scaling properties 
of multipartite entanglement for variable
system size, we now compute the time evolution of concurrence, precisely as 
in Fig.~\ref{decay3part}, for different values of $N$, and extract
the decay rates $\gamma$. Fig.~\ref{gammas}
shows the scaling of the entanglement decay rates $\gamma$ with $N$,
under the above, distinct decoherence mechanisms.

We see that the decay rate of 
the concurrence of GHZ states (top panel of
the figure) increases
linearly with $N$, except for the small-$N$ behaviour 
of $\gamma$ for the zero temperature environment. Indeed, in the
special case of dephasing environment, the density matrix is always a
mixture of two pure states, and, hence, of rank two, and can be treated
analytically. The observed behaviour comforts our intuition
-- which suggests that the larger the system, the easier it is for detrimental
environment effects to manifest. In addition, this fragility of GHZ
states was also observed in~\cite{briegel_PRL_2004} by an analysis based on
their separability and distillability properties~\cite{durPRL99,durPRA00} rather than a dynamical approach.

Remarkably, the situation changes quite drastically for the W states
(bottom plot of Fig.~\ref{gammas}). In this case, only the infinite
temperature environment gives rise to an almost linear increase of
$\gamma$ with $N$, slightly faster than for the GHZ states. In
contrast, for dephasing and zero temperature reservoirs, the decay of
the concurrence is {\em independent} of $N$. Moreover, the zero
temperature case also allows for an analytic solution (as above, the
rank of the state reduces to two) for all $N$, leading to $c_N(t)\sim
e^{-\Gamma t}$. Consequently, for these
environments, W states clearly
outperform 
GHZ states in terms of the robustness of their multipartite
entanglement properties.\footnote{Cluster states also present a kind
  of robust behaviour in terms of separability and distillability
  criteria in the case of a depolarising channel~\cite{briegel_PRL_2004}.}

One might be tempted to attribute this to the smaller initial concurrence
of W as compared to GHZ states (see Fig.~\ref{decay3part}), though the ratio
\begin{equation}
\frac{c_N(\Psi_{GHZ})}{c_N(\Psi_W)}=
\sqrt{(1-2^{1-N})\frac{N}{N-1}}\nonumber
\end{equation}
with a maximum at $N=5$, approaches unity for large $N$ (with a value $1.07$
for $N=7$, the maximal size considered in Fig.~\ref{gammas}).

\subsubsection{Experimental entanglement production: the M{\o}lmer-S{\o}rensen scheme \label{realist}}
In the previous sections
we assumed that
a perfect pure entangled state is available at the beginning, and
monitored the reservoir induced entanglement decay in the course of time.
Here, 
we will consider a situation where the
environment acts simultaneously to a unitary evolution which is
intended to {\em prepare} the entangled state. Specifically, 
we scrutinize a
scalable scheme proposed in~\cite{mol99} to produce GHZ-like states
(see Eq.~(\ref{ghzstate})),
\begin{equation}
\ket{\Psi_{N}}=\frac{1}{\sqrt{2}}\left ( \vert 00 \hdots 0
\rangle + e^{i\phi_N}\vert 11 \hdots 1 \rangle\right )\ ,
\label{msstates}
\end{equation}
experimentally implemented at NIST~\cite{wine00} to produce controlled
entanglement of 
two and four trapped ions.

\begin{figure}
\psfrag{Om0}{$\omega_0$}
\psfrag{n}{$\nu$}
\psfrag{ggn}{$\ket{0\,0\,\nu}$}
\psfrag{gen}{$\ket{0\,1\,\nu}$}
\psfrag{egn}{$\ket{1\,0\,\nu}$}
\psfrag{een}{$\ket{1\,1\,\nu}$}
\psfrag{gen+}{$\ket{1\,0\,\nu\hspace{-.1cm}+\hspace{-.1cm}1}$}
\psfrag{gen-}{$\ket{1\,0\,\nu\hspace{-.1cm}-\hspace{-.1cm}1}$}
\psfrag{egn+}{$\ket{0\,1\,\nu\hspace{-.1cm}+\hspace{-.1cm}1}$}
\psfrag{egn-}{$\ket{0\,1\,\nu\hspace{-.1cm}-\hspace{-.1cm}1}$}
\psfrag{D}{$\Delta$}
\epsfig{file=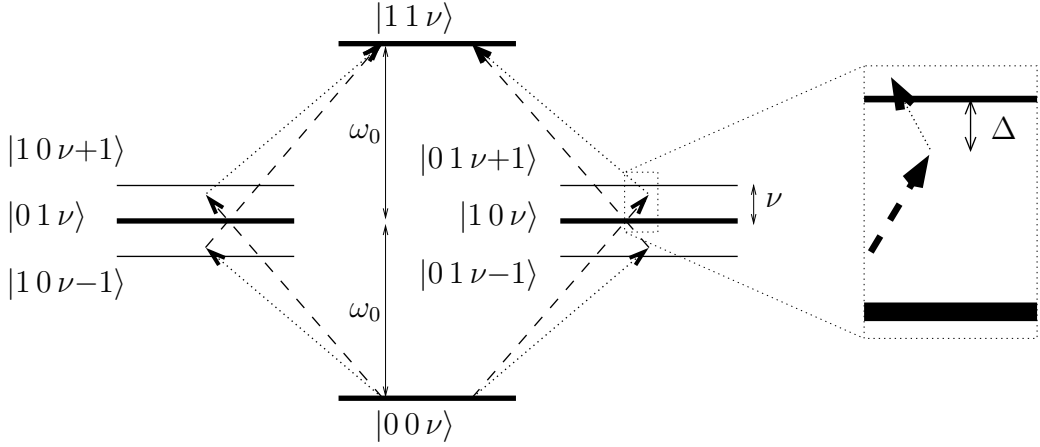,width=1.0\textwidth,angle=0}
\caption{Illustration of the M{\o}lmer-S{\o}rensen scheme \cite{mol99} for
the creation of a maximally entangled 
  state $(\ket{00}+i\ket{11})/\sqrt2$. Two ions are illuminated
  simultaneously by two electro-magnetic fields. One (dashed arrows)
  is red detuned with respect to the blue sideband, with detuning
  $\Delta$. The second pump field (dotted arrows) is blue detuned with
  respect to the red sideband. Both sideband transitions are not
  driven resonantly, however the two photon process
  $\ket{00}\leftrightarrow\ket{11}$
  is resonant, since the absolute values of the detunings of
  both transitions coincide.
  Thus, starting with the initial state $\ket{00}$,
  one can create a coherent superposition of the states
  $\ket{00}$ and $\ket{11}$.
}
\label{figms}
\end{figure}

To illustrate this preparation scheme -- which is 
valid for $N$ ions -- let us consider the case $N=2$ with two ions confined by
a harmonic potential,
simultaneously illuminated by two electromagnetic fields. The energy
levels are depicted in Fig.~(\ref{figms}), where $\omega_0$ is the
frequency of the electronic transition, and $\nu$ the oscillation
frequency of a given collective motional mode of the particles in the
trap. The fields oscillate with $\omega_0+ \nu - \Delta$ and $\omega_0
-\nu + \Delta$  in such a way that the two photon process that drives
the $\vert 00\rangle \leftrightarrow \vert 11\rangle$ transition is
resonant and a superposition of these states can be produced. At the
end of the preparation scheme, all one-photon processes, which excite
motional states, interfere destructively with each other, such that
vibrational and internal degrees of freedom are uncorrelated. However,
{\em during} the illumination these correlations do exist, and motional
decoherence can decrease the success probability of the preparation
process. The decay rate of the electronic states is negligible during
the time scale of the state preparation, and therefore one of the main
sources of errors in the system arises indirectly through the coupling with
these decohering motional modes. 

The heating of the ions, related to fluctuating
fields in the trap electrodes, leads to a thermal motion with steadily
increasing temperature and can be well described by
the infinite temperature reservoir discussed previously, Eq.~(\ref{TI}),
with measured heating rates $\Gamma/\nu$ ranging from $10^{-4}$ to
$10^{-3}$~\cite{winepra00,turch00}.  

\begin{figure}
\begin{center}
\includegraphics[width=10.0cm]{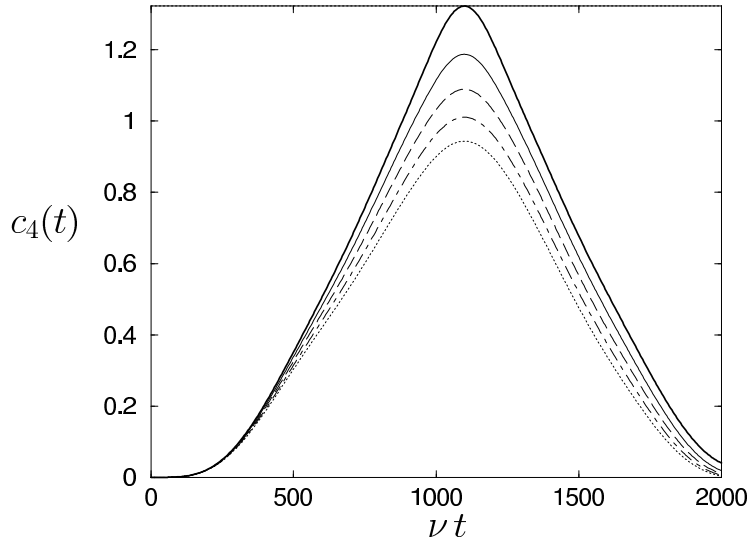}
\end{center}
\caption{Quasi-pure calculation (see Section~\ref{quasi-pure}) of
the time evolution of the 
  multipartite concurrence for four ions. In the case of vanishing environment
  coupling (bold 
  solid line), concurrence reaches its ideal value at the preparation time
  $\nu t_{\rm prep}= \pi/(\eta \Omega) \simeq 1100$, with $\eta=0.05$ and $\Omega=0.057$. As we increase 
the coupling strength $\Gamma/\nu$, the maximally achievable entanglement
  decreases, as shown for $\Gamma/\nu=1\times 10^{-4}$ (solid line),
  $2\times 10^{-4}$ (dashed line), $3\times 10^{-4}$
  (dot-dashed line) and $4\times 10^{-4}$ (dotted line).} 
\label{figms2}
\end{figure}

Figure~\ref{figms2} shows the time evolution of the multipartite
concurrence $c_N$, Eq.~(\ref{cN}), (evaluated through its quasi-pure approximation,
Eq.~(\ref{cqp})) under this scheme, for four ions 
and heating rates $\Gamma/\nu=0$ (bold solid line), $1\times 10^{-4}$
(solid line), $2\times 10^{-4}$ (dashed line), $3\times 10^{-4}$
(dot-dashed line) and $4\times 10^{-4}$ (dotted line). The system starts
with all ions in the electronic ground states and, during the
evolution, entanglement builds up until reaching its maximum value at
$\nu t_{prep} = \pi/(\eta \Omega) \simeq 1100$ (with $\eta$ the Lamb-Dicke
parameter, the ratio between the width of the vibrational ground
state and the wavelength of the driving radiation, and $\Omega$ the
single ion resonant Rabi frequency describing the strength of
ion-field coupling). After this preparation time the lasers should
be turned off and, for zero environment coupling, concurrence achieves
its ideal value corresponding to the GHZ-like state, while it shows appreciable decrease even for small heating rates. 

\begin{figure}
\begin{center}
\includegraphics[width=10.0cm]{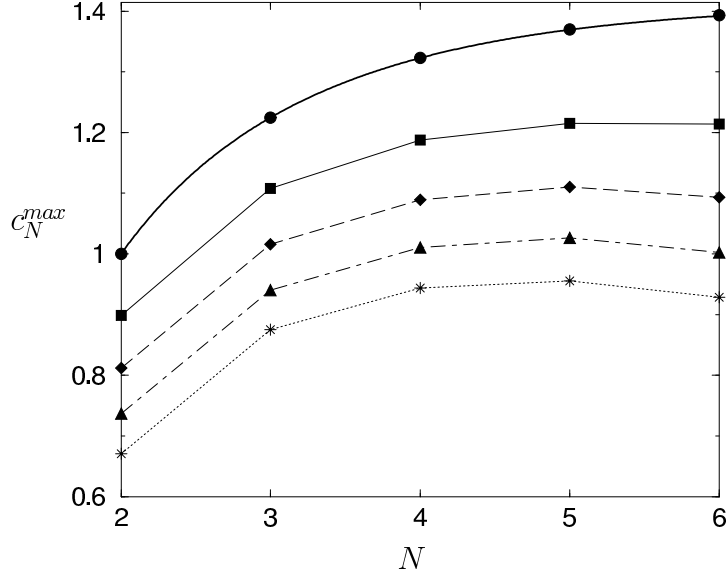}
\end{center}
\caption{Maximally achievable concurrence $c_N^{max}$, 
as a function of the number of ions, for reservoir coupling
  $\Gamma/\nu =0$ 
(circles), $1\times 10^{-4}$ (squares), 
$2\times 10^{-4}$ (diamonds), $3\times 10^{-4}$
  (triangles) and $4\times 10^{-4}$ (stars). The bold line shows the
  exact growth of the multipartite concurrence as a function of $N$
  for GHZ-like states, which saturates at $\sqrt{2}$. This growth is
  not monotonic anymore when motional decoherence is considered,
  indicating that environment effects become more and more detrimental
  for growing system size.} 
\label{figms3}
\end{figure}

\begin{figure}
\begin{center}
\includegraphics[width=10.0cm]{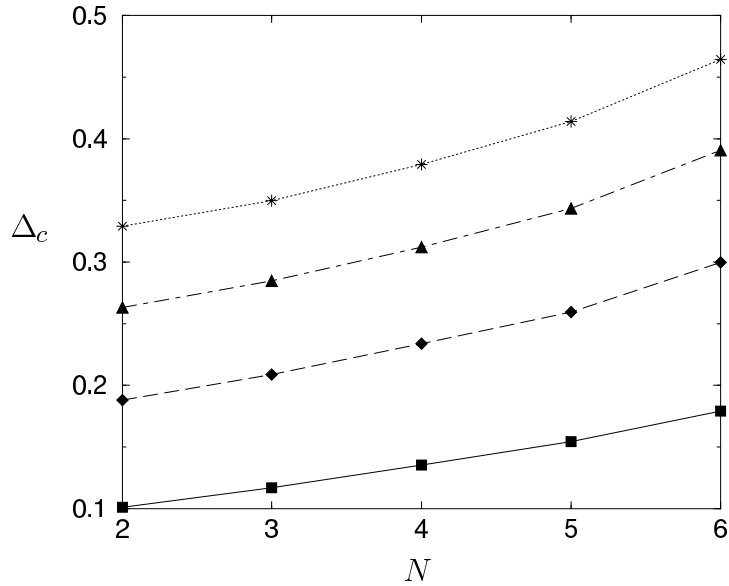}
\end{center}
\caption{Difference $\Delta_c=c_{N}^{\rm GHZ}-c_N^{\rm max}$ between the
  ideal and the maximally achievable 
concurrence, as a function
  of the system size $N$. Squares, diamonds, triangles and stars represent
reservoir coupling strengths 
  $\Gamma/\nu=1\times 10^{-4}$, $2\times 10^{-4}$, $3\times 10^{-4}$ and
  $4\times 10^{-4}$, respectively. Consistently with Fig.~\ref{figms3}, the 
entanglement loss
  during the preparation process increases with $N$. 
} 
\label{figms4}
\end{figure}

As in the case of the purely environment induced dynamics, we are
interested in the scaling of the achieved degree of entanglement with the
system size $N$. In the ideal
case, the maximum value of the multipartite concurrence, Eq.~(\ref{cN}), increases
with $N$ as $c_N=2^{1-N/2} 
\sqrt{(2^N-2)/2}$, saturating at $\sqrt{2}$ for large $N$ as
depicted by the bold solid line and circles in
Fig.~\ref{figms3}. With the addition of motional heating, this
growth with $N$ is not monotonic anymore and we observe, for example,
that from a certain value of the coupling strength the maximum concurrence for
six ions gets smaller than the one for five. Hence, 
the detrimental 
effect of (indirect) motional decoherence for entanglement
generation is enhanced 
with increasing system size.
This is further illustrated in Fig.~\ref{figms4}, where we plot the
entanglement loss $\Delta_c=c_N^{\rm GHZ}-c_N^{\rm max}$, i.e., the
difference between the ideal value and the actual optimal value achieved at a
finite decoherence level. 
For all values of the ion heating
rate, the entanglement loss increases with the system size.\footnote{Although
one might be tempted to associate this behaviour with the 
fragility of GHZ states with increasing number of particles as
discussed in Section~\ref{dynmultpart}, the connection is by no means
straightforward, since, in the present scheme, decoherence affects the
electronic levels only
indirectly.}

\section{Conclusion}
In the present contribution, we have performed a program reaching
from the formal definition to the time dependent monitoring 
of concurrence of arbitrary, finite
dimensional, multipartite quantum states under incoherent environment
coupling. The general applicability of the various lower
bounds of concurrence which have been derived stems from the algebraic
structure of a suitable reformulation of its original definition,
with direct generalizations for higher dimensional and/or multipartite
quantum states. 

We have checked the reliability and tightness of our various
approximate lower bounds of concurrence under very different physical
scenarios, from rather untypical entangled states with positive
partial transpose to experimental schemes of controlled environment
production. Whilst these results generally suggest a rather excellent
performance of the here derived, efficiently computable entanglement
measures, also quite a few questions remain to be answered on the
mathematical side: Which of our lower bounds of concurrence are
actually able to detect {\em all} entangled states? Are there
analytical error bounds for the various approximations? Are there
classes of entangled states which {\em cannot} be detected by our
lower bounds? Is there a suitable generalization of our
characterisation of multipartite entanglement through projectors on
antisymmetric subspaces such as to have a {\em complete} description
of multipartite quantum correlations? And is there a generalisation
for continuous variable systems?
 
And on the physical side: Given the now possible monitoring of the
time evolution of entanglement under environment coupling, how are the
entanglement decay rates encoded in the coherences of the original density
matrix? How do these decay rates depend on i) the initial state and
ii) the environment coupling operators and strengths? And,
perhaps most importantly: Since the Hilbert size dimension increases
exponentially with increasing systems size
(i.e. increasing number of system components), quantum state
tomography
will become inoperational to quantify the state's entanglement
properties. Which experimental observables are then best suited to
fulfill this task? Or, in other words, which are the robust dynamical  
observables which exhibit a clear 
experimental signature of the various types of multipartite entanglement? 

Hence, despite the considerable progresses on which we have reported
here, a panoply of challenging open questions awaits solution, what
simply reflects the potential -- on the fundamental level as well as
on the level of applications (from decoherence control to quantum
computation) -- of entanglement theory, a field still in its infancy, 
at the emerging interface of mathematical physics
and experiments.

\section{Acknowledgement}
We are indebted to
Rainer Blatt, Rafa{\l} Demkowicz-Dobrza{\'n}ski, Klaus Dietz, 
Berge Englert, Daniel
Est\`eve, Hartmut H\"affner, Peter H\"anggi, Armin Uhlmann, Thomas
Wellens, and Karol \.Zyczkowski, for inspiring
discussions, fruitful suggestions, critical comments and pertinent remarks.
Financial support by VolkswagenStiftung (under the project
``Entanglement measures and the influence of noise'') and Polish MNiI grant No
1P03B04226 is gratefully acknowledged.
This work was supported by a fellowship within the Postdoc-
Programme of the German Academic Exchange Service (DAAD).

\newpage

\bibliographystyle{elsart-num}

\bibliography{manuscript_physrep}

\end{document}